\documentclass[lettersize,journal]{IEEEtran}
\pdfoutput=1
\usepackage{amsmath,amsfonts}
\usepackage{algorithm}
\usepackage{algpseudocode}
\usepackage{array}
\usepackage{amssymb}
\usepackage[caption=false,font=normalsize,labelfont=sf,textfont=sf]{subfig}
\usepackage{textcomp}
\usepackage{stfloats}
\usepackage{url}
\usepackage{verbatim}
\usepackage{graphicx}
\usepackage{cite}
\usepackage{bm}
\usepackage{xcolor}
\usepackage{multirow}
\usepackage{epstopdf}
\usepackage{float}                  
\usepackage{subfig}                 
\usepackage{overpic}

\begin{document}

\title{Semantic Communications System with Model Division Multiple Access and Controllable Coding Rate for Point Cloud}

\author{Xiaoyi Liu,
        Haotai Liang,
        Zhicheng Bao,
        Chen Dong*,
        Xiaodong Xu,~\IEEEmembership{Senior Member,~IEEE.}
\thanks{Xiaoyi Liu, Haotai Liang, and Zhicheng Bao are with the State Key Laboratory of Networking and Switching Technology, Beijing University of Posts and Telecommunications, Beijing, China (e-mail: xiaoyiliu@bupt.edu.cn; lianghaotai@bupt.edu.cn; zhicheng\_bao@bupt.edu.cn).}
\thanks{*Chen Dong is the corresponding author and with the State Key Laboratory of Networking and Switching Technology, Beijing University of Posts and Telecommunications, Beijing, China (e-mail: dongchen@bupt.edu.cn).}
\thanks{Xiaodong Xu is with the State Key Laboratory of Networking and Switching Technology, Beijing University of Posts and Telecommunications, Beijing, China, and also with the Department of Broadband Communication, Peng Cheng Laboratory, Shenzhen, Guangdong, China (e-mail: xuxiaodong@bupt.edu.cn).
}
}

\markboth{Journal of \LaTeX\ Class Files,~Vol.~14, No.~8, August~2021}%
{Shell \MakeLowercase{\textit{et al.}}: A Sample Article Using IEEEtran.cls for IEEE Journals}

\maketitle

\begin{abstract}
Point cloud, as a 3D representation, is widely used in autonomous driving, virtual reality (VR), and augmented reality (AR). However, traditional communication systems think that the point cloud's semantic information is irrelevant to communication, which hinders the efficient transmission of point clouds in the era of artificial intelligence (AI). This paper proposes a point cloud based semantic communication system (PCSC), which uses AI-based encoding techniques to extract the semantic information of the point cloud and joint source-channel coding (JSCC) technology to overcome the distortion caused by noise channels and solve the ``cliff effect" in traditional communication. In addition, the system realizes the controllable coding rate without fine-tuning the network. The method analyzes the coded semantic vector's importance and discards semantically-unimportant information, thereby improving the transmission efficiency. Besides, PCSC and the recently proposed non-orthogonal model division multiple access (MDMA) technology are combined to design a point cloud MDMA transmission system (M-PCSC) for multi-user transmission. Relevant experimental results show that the proposed method outperforms the traditional method 10dB in the same channel bandwidth ratio under the PSNR D1 and PSNR D2 metrics. In terms of transmission, the proposed method can effectively solve the ``cliff effect" in the traditional methods.
\end{abstract}


\section{Introduction}
\IEEEPARstart{P}{oint} cloud is one of the representations of 3D, which uses geometric coordinates and other attributes (e.g., reflectance, etc.) to characterize points \cite{1}. Point cloud has been widely used in automatic driving, medical image processing, virtual reality (VR), and augmented reality (AR). Usually, the data volume of the point cloud is large, and the transmission of the point cloud will burden the traditional communication system, which does not pay attention to the meaning of the information to be transmitted and only encodes the information into strings with 0 and 1. For this reason, efficient transmission of the point cloud is highly desired.
  
According to Shannon's theory \cite{32_shannon1948mathematical}, communication is divided into three levels: symbol transmission, semantic exchange of transmitted symbols, and semantic information exchange effect. As a new communication paradigm, semantic communication system extracts the semantics of the information to be transmitted, encodes, and transmits it, improving communication efficiency. 

\begin{table*}[t]
\centering
\caption{semantic communication system for different sources.}
\label{table2}
\begin{tabular}{|l|l|l|l|}
\hline
Source & Model & Implemented function & OSI layer \\ \hline
Text & DeepSC & \begin{tabular}[c]{@{}l@{}}Recover the meaning of sentences by maximizing the system capacity \\ and minimizing the semantic errors.\end{tabular} & physical layer \\ \hline
\multirow{2}{*}{Image} & LSCI & \begin{tabular}[c]{@{}l@{}}Semantic slice-models (SeSM) are designed to represent the various mitigations of the \\ communication system to realize the flow of semantic intelligence.\end{tabular} & \multirow{2}{*}{physical layer} \\ \cline{2-3}
 & NTSCC & \begin{tabular}[c]{@{}l@{}}Incorporate hyperprior as side information into JSCC to realize \\ image-based semantic feature extraction and transmission.\end{tabular} &  \\ \hline
Video & DVST & \begin{tabular}[c]{@{}l@{}}Map video sources into nonlinear potential space to achieve \\ video-based semantic feature extraction and transmission.\end{tabular} & physical layer \\ \hline
\multirow{3}{*}{3D point cloud} & AITransfer & \begin{tabular}[c]{@{}l@{}}Incorporate the dynamic network condition into an end-to-end \\ encoder-decoder network to provide an adaptive transmission control scheme.\end{tabular} & \multirow{2}{*}{network layer} \\ \cline{2-3}
 & IsCom & \begin{tabular}[c]{@{}l@{}}Design an ROI selection module, a lightweight encoder-decoder network, \\ and an intelligent schedular for online adaptive point cloud video service.\end{tabular} &  \\ \cline{2-4} 
 & PCSC & \begin{tabular}[c]{@{}l@{}}Design a point cloud semantic communication system and realizes the \\ controllable coding rate without fine-tuning the network.\end{tabular} & physical layer \\ \hline
\end{tabular}
\end{table*}

\par Semantic communication processes data in the semantic domain by extracting the meaning of the data and filtering out unimportant information, compressing the data while preserving the meaning. In semantic communication, a new joint source-channel coding (JSCC) scheme based on deep neural network (DNN) is presented \cite{23_Deepjscc-q}\cite{24_Deep_joint_source-channel}\cite{25_DeepJSCC-f}\cite{34-neural_joint_source_channel_coding}\cite{33_Deepjscc}. This scheme is robust to the harsh channel environment, i.e., the low signal-to-noise ratio (SNR) region, and can solve the ``cliff effect" well. There are already some semantic communication systems, as shown in Table \ref{table2}. For text communication system, DeepSC \cite{16_DeepSC} maximizes system capacity and minimizes semantic errors by restoring the meaning of a sentence rather than using bit or symbol errors in conventional communication. For image communication systems, the NTSCC \cite{17_NTSCC} uses JSCC based on nonlinear transformation to realize image-based semantic feature extraction and transmission. LSCI \cite{26_LSCI} believes that the semantic transmission is essentially the flow of the AI model, so the semantic slice model (SeSM) is designed to realize this idea. DVST \cite{20_DVST} also uses a nonlinear transformation to achieve semantic video transmission. The semantic communication systems mentioned above are all based on the physical layer of the open system interconnect reference model (OSI). There are also studies on network layer semantic communication systems based on point cloud video \cite{18_AITransfer}\cite{19_ISCom}. In AITransfer \cite{18_AITransfer}, the dynamic network condition is incorporated into the end-to-end point cloud compression architecture. It employs a deep reinforcement learning-based adaptive control scheme to provide robust transmission. ISCom \cite{19_ISCom} consists of a region-of-interest (ROI) selection module, a lightweight point cloud video encoder-decoder network, and a deep reinforcement learning (DRL)-based scheduler to adapt an optimal encoder-decoder network. 
\par However, there is still room for improvement. First,  as mentioned above, the current point cloud semantic communication system does not involve the research of the channel in the physical layer. Therefore, it is necessary to design a physical-layer-based point cloud semantic communication system. In this paper, a point cloud semantic communication system, named PCSC, is proposed.
\par Second, in semantic communication systems, many methods cannot manually and accurately control code lengths in a simple way. For example, in \cite{17_NTSCC}, the code length is learned and controlled implicitly, and it is not easy to give an accurate code length at a fixed channel bandwidth ratio. In \cite{16_DeepSC}, the specific network parameters need to be changed and retrained to get different coding rates, which consumes lots of computing power and storage space. A rate allocation network is introduced in \cite{21_zhang} to estimate the code length, although it does not need to be retrained to get different coding rates, the trained rate allocation network takes up storage space. Therefore, we propose three methods to analyze the importance of the encoded semantic vector and discard the non-important vector according to the specified bandwidth ratio to achieve efficient transmission. In this process, the network does not need to fine-tune for a particular coding rate. 
\par Third, according to the recently proposed non-orthogonal model division multiple access (MDMA) \cite{22_MDMA}, semantic features extracted from the same artificial intelligence model (AI) have some shared information and some personalized information. In this paper, the non-orthogonal MDMA and point cloud semantic transmission are combined to construct a point-cloud-based MDMA transmission system, named M-PCSC, validating that non-orthogonal MDMA is also applicable to sources in other modalities. 
\par The contribution of this paper can be summarized as the following:
\par (1) PCSC framework: A novel end-to-end learnable framework for point cloud transmission, i.e., PCSC, is proposed, which can extract the point cloud semantic information effectively.  A joint source-channel coding (JSCC) is exploited to cope with channel noise and semantic distortion, which leads to a more robust point cloud transmission than traditional transmission schemes. To the best of the authors' knowledge, this is the first paper to propose a point cloud semantic communication system in the physical layer.
\par (2) Rate controllable semantic feature transmission: This paper uses a simple method to achieve different coding rates without any other rate estimate models or complicated frameworks. The PCSC, which is trained only once, can generate a wide range of variable coding rates. This mechanism is achieved by the value and the grad value of the encoded semantic vector in the latent representation.
\par (3) MDMA transmission: This paper presents a point cloud semantic communication system based on non-orthogonal MDMA (M-PCSC) to realize multi-user transmission, which can save channel bandwidth. We also found that compression is mainly aimed at shared information. As the compression rate decreases, the amount of shared information will reduce, and only the personalized information is left. Besides, the semantic spectral efficiency (S-SE) of downlink M-PCSC is deduced, and the S-SE is optimized to obtain the maximum spectral efficiency.
\par (4) Performance validation: We verify the performance of the PCSC across standard datasets. In the same channel bandwidth ratio (CBR), the PCSC has a much better gain on various established metrics, such as PSNR D1 and PSNR D2, compared to traditional G-PCC/Point Cloud Library (PCL) combined with LDPC and digital modulation schemes. Experiments show that the PCSC outperforms the traditional methods 10dB average in the same channel bandwidth ratio under the PSNR D1 and PSNR D2 metrics. 
\par The rest of this paper is arranged as follows: In Section \uppercase\expandafter{\romannumeral2}, the overall system model of PCSC and M-PCSC is introduced. Then, introduce the network architecture and the algorithms in detail in Section \uppercase\expandafter{\romannumeral3}. Section \uppercase\expandafter{\romannumeral4} shows the experiment setup and results. Finally, the paper is concluded in Section \uppercase\expandafter{\romannumeral5}.

\begin{figure}[b]
\centering
\includegraphics[width=\linewidth]{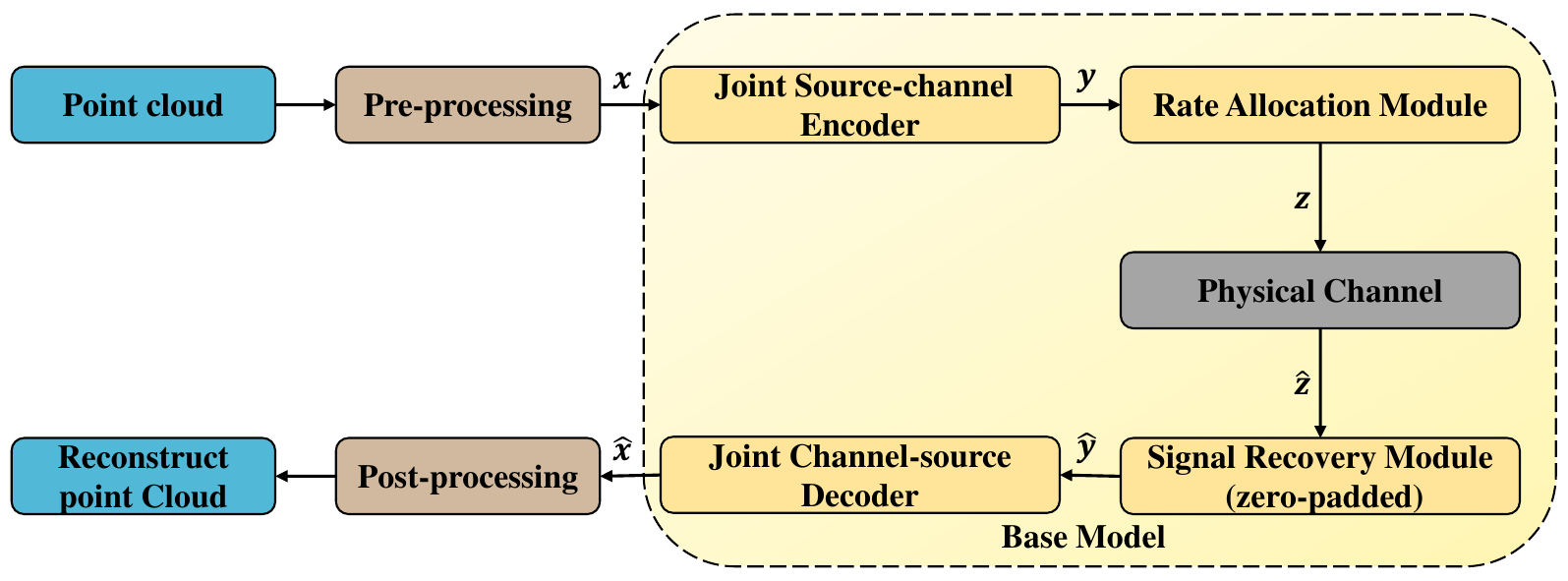}
\caption{The framework of semantic communication system, PCSC. The transmitter includes a joint source-channel encoder and a rate allocation module. The transmitter encodes the pre-processed point cloud $x$ into $y$ and discards some semantically-unimportant information to get the shortened vector $z$. The receiver includes a signal recovery module and a joint channel-source decoder. The signal recovery module fills $\hat{z}$ with zero to ensure that the length of $\hat{y}$ is equal to $y$, and then the decoder decodes $\hat{y}$ to get $\hat{x}$. Finally, post-processing the $\hat{x}$ to obtain the reconstructed point cloud.}
\label{fig1}
\end{figure}

\section{System Model}
This section describes the proposed point cloud semantic communication system. First, an end-to-end point cloud semantic communication system (PCSC) with a stochastic physical channel is introduced. Then, a point cloud semantic communication system with non-orthogonal model division multiple access (M-PCSC) is presented.

\begin{figure}[b]
\centering
\includegraphics[width=\linewidth]{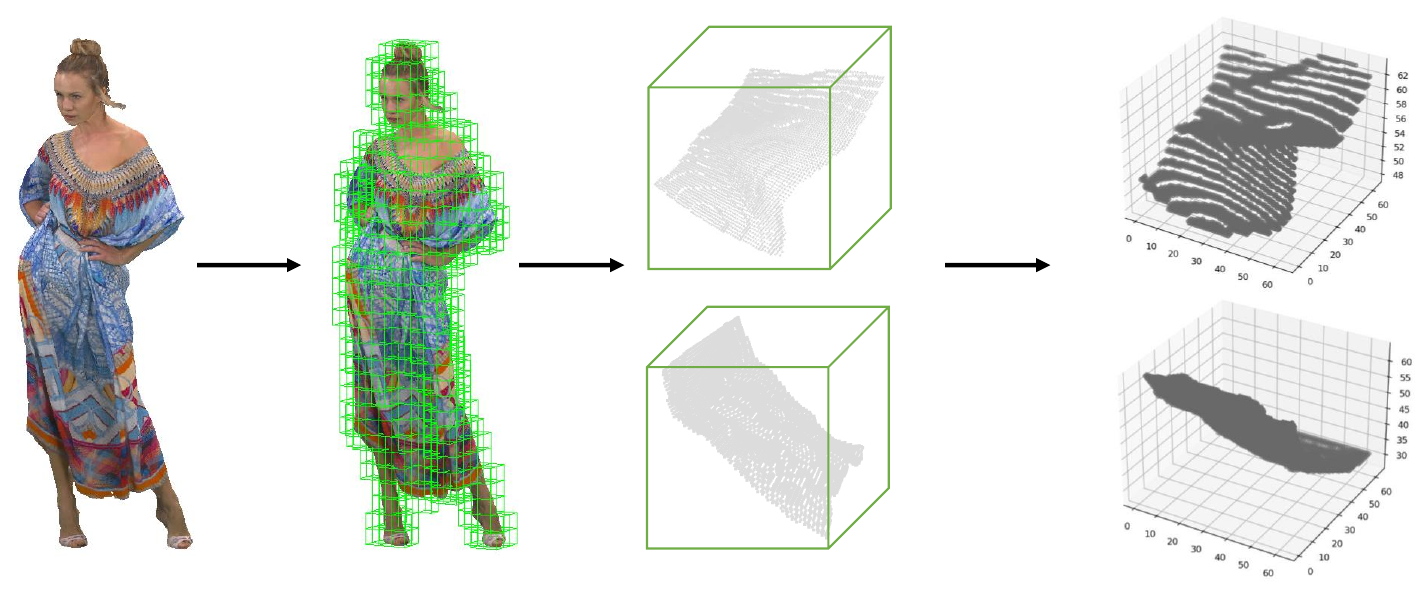}
\caption{Point cloud partitioning: the point cloud is divided into cubes and then is represented as $W\times W\times W$ binary voxels.}
\label{fig2}
\end{figure}

\subsection{End-to-End Point Cloud Semantic Communication System}
The overall architecture of the proposed PCSC is presented in Fig. \ref{fig1}. After preprocessing the point cloud to get $x$, the joint source-channel encoder map $x$ into a semantic vector $y$. The rate allocation module adopts a particular approach to rank the semantic vector $y$  based on its level of importance, discarding non-essential information within $y$ according to the restrictions, which can be set manually and explicitly. This selective process enables variable length coding to be executed on $y$, converting $y$ into $z$. Term $z$ is passed through the physical channel with transmission impairments, such as noise and distortion. The received, $\hat{z}$, is restored to $\hat{y}$ in the signal recovery module by zero-padding. The $\hat{y}$ is decoded into $\hat{x}$ at the joint channel-source decoder and recovered to the point cloud after the post-processing. Both the encoder and decoder are designed with DNNs and trained together. The various PCSC modules are described in detail as follows:

\subsubsection{Pre-processing}
A point cloud is composed of a large volume of voxels when represented using ($i$, $j$, $k$)-based Manhattan space. The computational coding complexity grows significantly if processing an entire point cloud at a time, especially for a point cloud with high precision. For example, a point cloud with 10-bit precision allows $0\leq i, j, k \leq 2^{10}-1$. Thus, the point cloud is partitioned into non-overlap cubes, as shown in Fig. \ref{fig2}, to reduce computational complexity. The coordinates of each cube are described using the octree decomposition method \cite{1}. Assuming that the size of the cube is \emph{W×W×W}, and the position of a particular cube is ($i_{c}$,$j_{c}$,$k_{c}$), the local coordinates of a voxel ($i_{l}$,$j_{l}$,$k_{l}$) can be represented using global coordinate ($i_{g}$,$j_{g}$,$k_{g}$):
\begin{equation}
\label{eq1}
(i_{l},j_{l},k_{l}) = (i_{g},j_{g},k_{g})-W\times(i_{c},j_{c},k_{c}).
\end{equation}

\subsubsection{Base model}
As shown in Fig. \ref{fig1}, the transmitter consists of a joint source-channel encoder and a rate allocation module. The joint source-channel encoder extracts the semantic features from partitioned point cloud $x$ and ensures the effective transmission of semantic information over the physical channel simultaneously. The encoded symbol stream can be represented as follows:
\begin{equation}
\label{eq2}
y=SC{_{\alpha}}(x),
\end{equation}
where $SC{_{\alpha }}(\cdot)$ is the joint source-channel encoder with parameter ${\alpha}$.

\par The rate allocation module discards some points according to the importance degree of the semantic information to achieve variable rate coding and generates the semantic vector $z$:
\begin{equation}
\label{eq3}
z=R(y),
\end{equation}
where $R(\cdot)$ is the rate allocate function, and the method for ranking the importance of the semantic factor is given in Section \uppercase\expandafter{\romannumeral3}.
\par After encoding, the shortened vector $z$ will be transmitted through the wireless channel. The received $\hat{z}$ at the receiver can be expressed as:
\begin{equation}
\label{eq4}
\hat{z}=h\times z+n,
\end{equation}
where $h$ corresponds the Rayleigh fading channel with $\mathcal{CN}$(0,1), $n$ is the additive white Gaussian noise with $\mathcal{CN}$(0,$\sigma ^{2}$). For the additive-white-Gaussian-noise-channel (AWGN), $h$=1. For the end-to-end training of the encoder and the decoder, the channel must allow for backpropagation. In this paper, for the sake of simplicity, the AWGN and the Rayleigh fading channel are mainly considered.
\par At the receiver, the signal recovery module performs zero padding on the received $\hat{z}$ to ensure that the lengths of the padded semantic vector $\hat{y}$ and $y$ are equal. The padded semantic vector $\hat{y}$ can be expressed as:
\begin{equation}
\label{eq5}
\hat{y}=Re(\hat{z}),
\end{equation}
where $Re(\cdot)$ is the signal recovery function. The recovered $\hat{y}$ is sent to the joint channel-source decoder for decoding, which can be represented as:
\begin{equation}
\label{eq6}
\hat{x}=SC_{\beta }^{-1}(\hat{y}),
\end{equation}
where $SC_{\beta }^{-1}(\cdot)$ is the joint channel-source decoder with parameter $\beta$.
\par The PCSC aims to decrease semantic errors and minimize the number of transmitted symbols. Despite the ability of current communication systems to accomplish transmission with a low bit error rate, the presence of even a few bit errors may result in significant distortion of the reconstructed point cloud due to channel noise. This occurs due to the absence of certain point cloud information. To recover the point cloud successfully at the semantic level, this paper employs the joint source-channel coding to keep the meaning between $x$ and $\hat{x}$ unchanged. In the pre-processing, 1 indicates that the voxel is occupied, and 0 suggests that the voxel is not occupied. The decoded point cloud is a floating number in the range of 0 to 1, so the decoded point cloud needs to be classified into 1 or 0 accordingly. Inspired by \cite{5Lossy}, we use weighted binary cross entropy (WBCE) as the distortion in training, and the formula is as follows:
\begin{equation}
\label{eq7}
l_{WBCE}=\frac{1}{N_{o}}\sum^{N_{o}}-\log p_{\widetilde{x}_o}+\zeta \frac{1}{N_{n}}\sum^{N_{n}}-\log (1-p_{\widetilde{x}_n}),
\end{equation}
where $p_{\widetilde{x}}=sigmoid(\widehat{x})$, which is used to estimate the probability of a voxel being occupied, $\widetilde{x}_o$ denotes occupied voxels, $\widetilde{x}_n$ denotes unoccupied voxels, $N_{o}$ and $N_{n}$ denotes the number of occupied and unoccupied voxels, respectively. The value of the voxel $\widetilde{x}$ is a floating number between 0 and 1, which guarantees the differentiability of the backpropagation. At the same time, $\zeta$ is used to calculate the average loss of positive and negative samples to balance the loss penalty. In the experiment, $\zeta$=3.

\subsubsection{Post-processing} In the post-processing, the voxels decoded by the joint channel-source decoder are floating numbers of 0 to 1. They need to be binarized, using only 0 and 1 to represent the voxels. This paper uses the adaptive threshold for binarization \cite{5Lossy}. Since $p_{\widetilde{x}}$ can also be considered as the probability of being occupied, $p_{\widetilde{x}}$  is sorted to extract the first $k$ voxels that are most likely to be occupied. The binarized points are converted into local coordinates $(i_{l},j_{l},k_{l})$, then converted into global coordinates $(i_{g},j_{g},k_{g})$ and finally merged into a complete point cloud:
\begin{equation}
\label{eq8}
(i_{g},j_{g},k_{g})=(i_{l},j_{l},k_{l})+W\times (i_{c},j_{c},k_{c}),
\end{equation}
where $(i_{c},j_{c},k_{c})$ is the position of a particular cube.

\subsection{Non-orthogonal Model Division Multiple Access for Point Cloud Communication System}
A new type of non-orthogonal multiple access technology based on semantic domain resources, named model division multiple access (MDMA), has been proposed recently \cite{22_MDMA}. When multi-user transmission is performed, to save transmission bandwidth, the shared information is transmitted only once, and the personalized information of each user is transmitted separately. This paper designs a non-orthogonal MDMA transmission system based on point clouds named M-PCSC. The overall architecture is shown in Fig. \ref{fig.3}.

\subsubsection{Uplink System}
\begin{figure}[t]
	\centering
	\subfloat[]{\includegraphics[width=\linewidth]{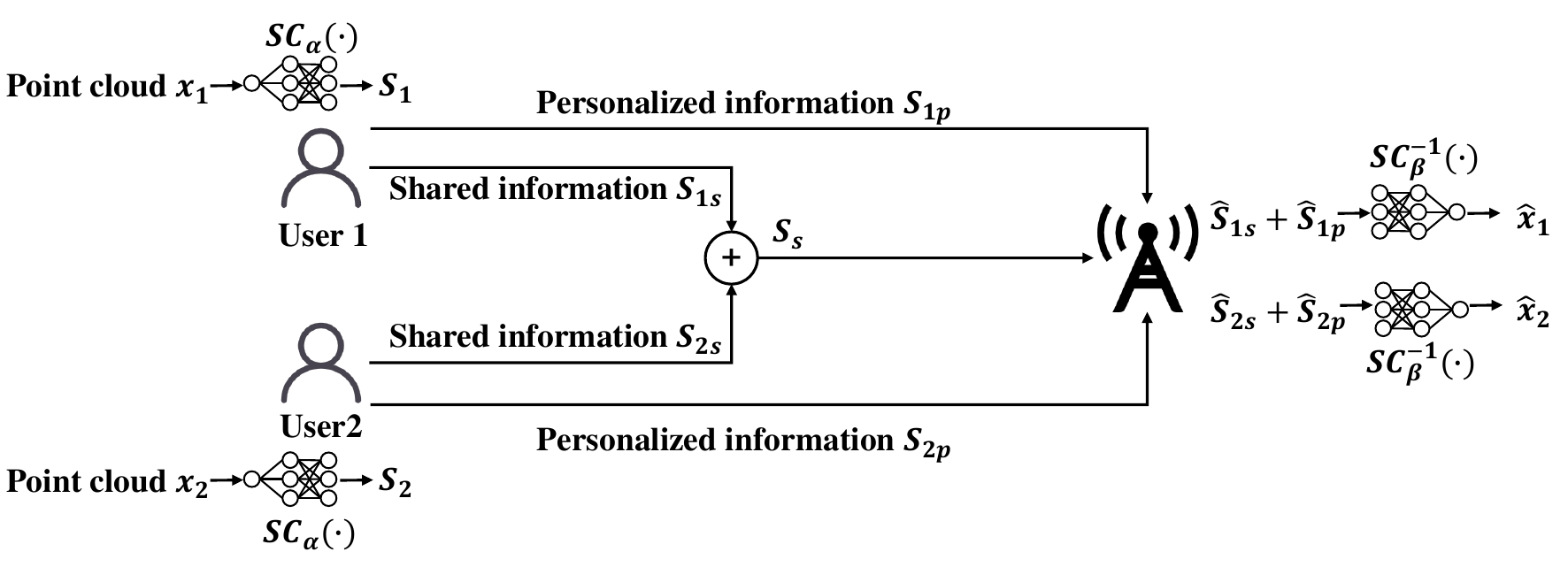}\label{fig3a}}\
    \subfloat[]{\includegraphics[width=\linewidth]{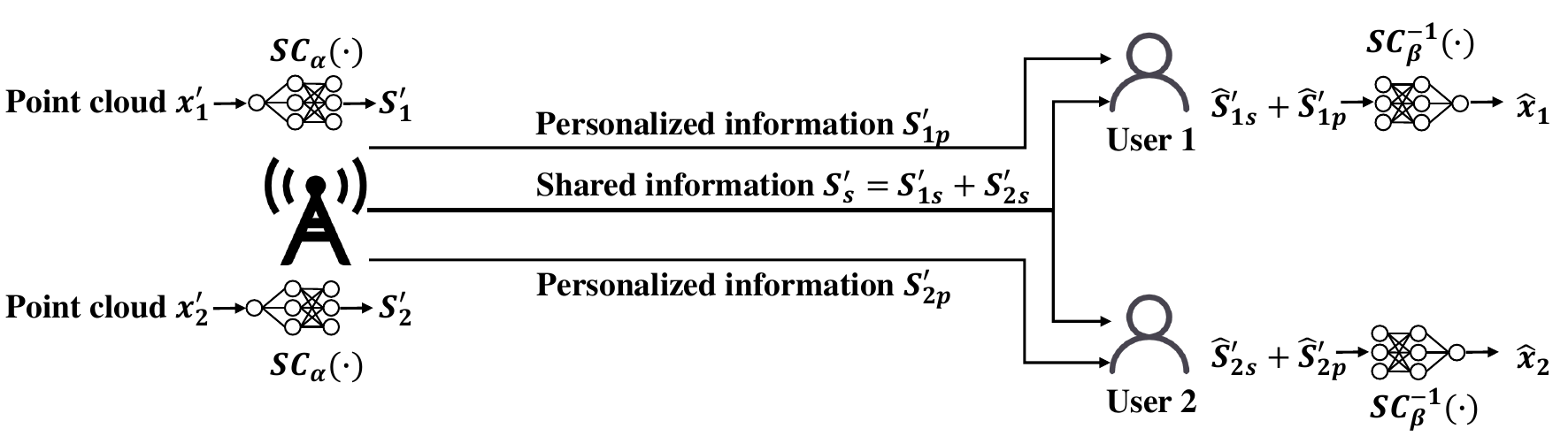}\label{fig3b}}\
	\caption{The framework of M-PCSC. (a): the uplink framework of M-PCSC.  (b) the downlink framework of M-PCSC.}
    \label{fig.3}
\end{figure}

The overall framework of the uplink transmission system is shown in Fig. \ref{fig.3}\subref{fig3a}. First, user 1 and user 2 extract the semantic information $S_{1}$ and $S_{2}$ by using the joint source-channel encoder:
\begin{equation}
\label{eq9}
S_{1}=SC_{\alpha}(x_{1}), S_{2}=SC_{\alpha}(x_{2}),
\end{equation}
where $SC_{\alpha}(\cdot)$ is the joint source-channel encoder in the PCSC with parameter $\alpha$. The shared information $S_{1s}$, $S_{2s}$ and the personalized semantic information $S_{1p}$, $S_{2p}$ can be obtained according to the specific agreements. At time slot 1 (frequency 1), the shared semantic information $S_{1s}$ and $S_{2s}$ is merged at the air interface as $S_s$ and sent through the wireless channel. At time slot 2 (frequency 2), the personalized semantic information $S_{1p}$ and $S_{2p}$ are sent to the base station. The received $\hat{S}_{s}$, $\hat{S}_{1p}$, and $\hat{S}_{2p}$ can be represented as follows:

\begin{equation}
\label{eq25}
\left\{
             \begin{array}{lr}
             \hat{S}_{s}=S_s+n_0, &  \\
             \hat{S}_{1p}=S_{1p}+n_1, & \\
             \hat{S}_{2p}=S_{2p}+n_2, &  
             \end{array}
\right.
\end{equation}
where $n_0$, $n_1$, $n_2$ is the additive white Gaussian noise with $CN(0, \sigma^{2})$ when transmitting the shared information $S_s$, and personalized information $S_{1p}$, $S_{2p}$, respectively. After receiving $\hat{S}_{s}$, $\hat{S}_{1p}$, and $\hat{S}_{2p}$, the base station first extracts $\hat{S}_{1s}$ and $\hat{S}_{2s}$ from $\hat{S}_{s}$ using function $f(\cdot)$:
\begin{equation}
    \label{eq27}
    \hat{S}_{1s}=f(\hat{S}_{s}), \hat{S}_{2s}=f(\hat{S}_{s}),
\end{equation}
Then the base station can directly store the semantic information of $x_1$ and $x_2$, or recover the original point cloud $\widehat{x}_{1}$, $\widehat{x}_{2}$ by using the joint channel-source decoder:
\begin{equation}
    \label{eq10}
    \hat{x}_1=SC_{\beta }^{-1}(\hat{S}_{1s}+\hat{S}_{1p}), \hat{x}_2=SC_{\beta }^{-1}(\hat{S}_{2s}+\hat{S}_{2p}),
\end{equation}
where $SC_{\beta }^{-1}(\cdot)$ is the joint channel-source decoder in the PCSC with parameter $\beta$.

\subsubsection{Downlink System}
The overall framework of the downlink system is shown in Fig. \ref{fig.3}\subref{fig3b}. First, the base station extracts the semantic information ${S'_{1}}$ and ${S'_{2}}$ of the point cloud ${x'_{1}}$ and ${x'_{2}}$ using the joint source-channel joint encoder $SC_{\alpha}(\cdot)$:

\begin{equation}
    \label{eq11}
    {S'_{1}}=SC_{\alpha}({x'_{1}}), {S'_{2}}=SC_{\alpha}({x'_{2}}).
\end{equation}
The shared information ${S'_{1s}}$, ${S'_{2s}}$ and the personalized information ${S'_{1p}}$, ${S'_{2p}}$ can be extracted by comparing the absolute variance between ${S'_{1}}$ and ${S'_{2}}$. The shared information ${S'_{s}}$ superimposed ${S_{1s}}'$ and ${S_{2s}}'$ is sent only once, and the personalized information ${S'_{1p}}$ and ${S'_{2p}}$ are transmitted respectively to the users through the wireless channel: 

\begin{equation}
\label{eq26}
\left\{
             \begin{array}{lr}
             \hat{S}^{'}_{s}=S'_{s}+n'_0, &  \\
             \hat{S}^{'}_{1p}=S'_{1p}+n'_1, & \\
             \hat{S}^{'}_{2p} =S'_{2p}+n'_2, &  
             \end{array}
\right.
\end{equation}
where $n'_0$, $n'_1$, $n'_2$ is the additive white Gaussian noise with $CN\left(0,\sigma^2\right)$. The $\hat{S}^{'}_{1s}$ and $\hat{S}^{'}_{2s}$ can be obtained using the $f(\cdot)$ mentioned in Eq. (\ref{eq27}). Then user1 and user2 utilize the joint channel-source decoder to reconstruct the point cloud:
\begin{equation}
    \label{eq12}
    {\hat{x}^{'}_{1}}=SC_{\beta}^{-1}(\hat{S}^{'}_{1s}+\hat{S}^{'}_{1p}), {\hat{x}_{2}}'=SC_{\beta}^{-1}(\hat{S}^{'}_{2s}+\hat{S}^{'}_{2p}).
\end{equation}

\section{System Design}
\begin{figure*}[t]
\centering
\includegraphics[width=\linewidth]{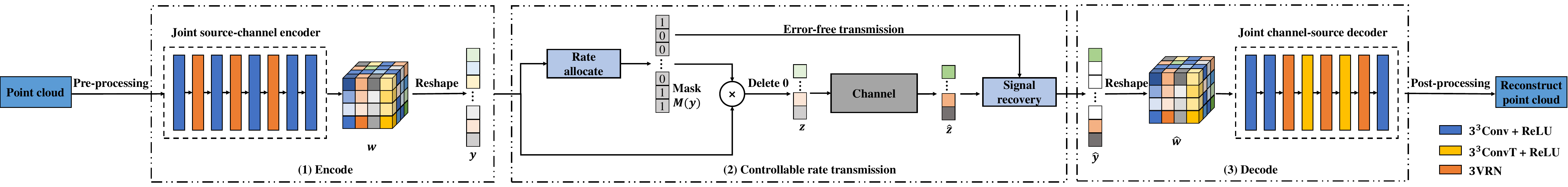}
\caption{The design of the PCSC. The transmitter includes a joint-source channel encoder and a rate allocation module to rank the semantic vector and discard some semantically-unimportant vectors. The receiver includes a signal recovery module to conduct zero-padding to the received shortened vector and a joint channel-source decoder.}
\label{fig4}
\end{figure*}

This section mainly introduces the specific design of the PCSC and the M-PCSC. For the PCSC, as shown in Fig. \ref{fig4}, the transmitter includes a joint source-channel encoder and a rate allocation module. It extracts semantic information from the point cloud to be transmitted and generates symbols to facilitate subsequent transmission. The receiver mainly includes a signal recovery module and a joint channel-source decoder. This section first introduces the design of the PCSC and then introduces the controllable rate coding method. Finally, as shown in Fig. \ref{MDMA_protocol}, the design of the M-PCSC is introduced.

\subsection{Base Model}
The base model is divided into three parts, as shown in Fig. \ref{fig4}. Each part will be described in detail as follows:
\subsubsection{Encode}
The pre-processed point cloud is fed into the joint source-channel encoder, which is based on 3D convolution. Stacking 3D convolutional neural networks makes sparsely distributed occupied voxels more compact. Voxception-Resnet (VRN) \cite{7VRN} is the basic unit of the joint source-channel encoder and the joint channel-source decoder. The VRN structure is based on the residual network structure \cite{27_Resnet}, but the bottleneck block in the residual structure and the basic residual network are connected into a single Inception-style block. Relevant research \cite{5Lossy} has used VRN in the point cloud hyperprior coding structure. The encoder and decoder architecture used in this paper refers to the network architecture proposed in \cite{5Lossy} and \cite{7VRN}. The joint source-channel encoder consists of an initial convolution layer, three main units, and two convolutions at the end. Each unit contains three stacked VRN blocks and a downsample convolutional layer. The encoder takes $x$ as the input and reshapes the output $w$ as a one-dimensional vector $y$. 
\subsubsection{Controllable rate transmission}
The rate allocation module will generate a 0-1 mask, $M(y)$, according to the importance ranking of $y$ and the restrictions, which can be set manually and explicitly. The importance ranking detail will be given later. Term $M(y)$ is element-wisely multiplied with $y$, and the results with 0 values will be discarded to get the transmitted symbols $z$. The shortened symbol $z$ is transmitted through a wireless channel, and the $M(y)$ will be transmitted to the receiver error-free. After receiving the $\hat{z}$, in the signal recovery module, zero-padding is applied to $\hat{z}$ to ensure the length of $\hat{y}$ equals $y$. 
\subsubsection{Decode}
Term $\hat{y}$ is reshaped into $\hat{w}$ in the same form as $w$. After this, $\hat{w}$ will be fed into the joint channel-source decoder, which has a symmetric architecture with the joint source-channel encoder. 
\par Finally, the reconstructed point cloud can be obtained by post-processing the decoded result $\hat{x}$.

\begin{algorithm}[t]
\caption{Controllable coding rate algorithm.}\label{alg:cap}
\begin{algorithmic}[1]
\Require pre-processed point cloud $x$, discard points ratio, method
\State\textbf{Transmitter:}
\State$w \gets SC_{\alpha}(x)$
\State$y \gets$ reshape $w$
\State points num $\gets$ len($y$) $\times$ drop rates ratio
\If{method is ``value"}
\State index $\gets$ argsort($\left | y\right |$)[points num:]
\ElsIf{method is "grad" or ``grad $\times$ value"}
\State $x \gets SC_{\beta}^{-1}(y)$
\State Compute loss by Eq. (\ref{eq7})
\State $y_{grad}$ $\gets$ loss.backward() 
\If{method is ``grad"}
\State index $\gets$ argsort($\left |y_{grad}\right |$)[points num:]
\ElsIf{method is ``grad $\times$ value"}
\State index $\gets$ argsort($\left |y_{grad} \times y\right |$)[points num:]
\EndIf
\EndIf
\State mask $\gets$ 1
\State mask[index] $\gets$ 0
\State $z$ $\gets$ $y$ $\times$ mask
\State Delete zeros in $z$.
\State Transmit $z$, transmit mask error-free.
\State \textbf{Receiver:}
\State Receive $\hat{z}$, mask.
\State $cnt $$\gets$ 0, $i$ $\gets$ 0
\While{$i$$<$len(mask)}
\If{$mask[i]$$==$0}
\State$\hat{y}[i]\gets$0
\ElsIf{$mask[i]$$\neq$0}
\State $\hat{y}[i]\gets \hat{z}$[cnt]
\State $cnt\gets cnt+$1
\EndIf
\State$ i\gets i+1$
\EndWhile
\State$\hat{w} \gets$ reshape $\hat{y}$
\State$\hat{x} \gets SC_{\beta}^{-1}$($\hat{z}_{padding}$)
\Ensure $\hat{x}$
\end{algorithmic}
\end{algorithm}
\subsection{Controllable Coding Rate}
A rate allocation module is introduced in this paper to encode the point cloud with different rates. The methods do not need to train a specific rate allocation network or modify the original network parameters for re-training. They also can generate arbitrary coding rates (less than the maximum coding rate of the system). It can also be used in other semantic communication systems to achieve variable rate coding, which does not change the original structure. This paper uses the channel bandwidth ratio (CBR) \cite{25_DeepJSCC-f} to measure the coding rate. $CBR=\frac{k}{e} (k<e)$, where $e$ is the source bandwidth given by the product of the pre-processed point cloud cube's length, width, height, and number of channels. Term $k$ is defined as the channel bandwidth, and the value of $k$ is the total length of the semantic vector for transmission. This subsection mainly describes how to analyze the importance of the semantic vector after coding by the encoder.

\subsubsection{Value of the semantic vector}
The semantic vector $y$ can be expressed by $[1,L]$, where $L$ represents the length of the semantic vector. For each vector $y_{i} (i<L)$, ${\left | y_i\right |}^2$ represents the signal power. So, a larger ${\left | y_i\right |}$ represents a more significant power. During the training of the PCSC, the network will allocate high power for vectors of great importance, and this point can be proved in subsequent experiments. For this reason, it can be considered that a larger ${\left | y_i\right |}$ has a higher degree of importance and a more robust noise tolerance. The network will rely more on it for subsequent decoding. More minor ${\left | y_i\right |}$ are values that the network considers unimportant after learning, and the network can still decode well even if these insignificant points are not transmitted to the receiver. Therefore, the absolute values of the encoded semantic vectors can be ordered, and signals with small absolute values, i.e., signals with small power, can be discarded in the process of signal transmission to save channel bandwidth.
\subsubsection{Gradient of the semantic vector}
For the PCSC, during the training process, according to Eq. (\ref{eq7}), both the $\hat{x}$ recovered from the decoder and the input $x$ are used to calculate the loss, the loss is back-propagated to update the network weight. 
The gradient of the encoded semantic vector $w$ can be obtained in the backpropagation of the loss function. Assuming that the dimension of the encoded semantic vector is $[A,C,E,F,G]$, the gradient $\delta^{p}$ of the $p^{th}$ channel in the semantic vector is as follows:
\begin{equation}
    \label{eq13}
    \delta ^{p}=\frac{1}{E\times F \times G}\sum_{i}\sum_{j}\sum_{k}\frac{\partial l}{\partial w_{ijk}^{p}},
\end{equation}
where $l$ is the loss function, $w_{ijk}^p$ represents the value of $w$ in the channel $p$ at the coordinates $(i,j,k)$. The gradient value $\delta^p$ of each channel in $w$ is calculated using Eq. (\ref{eq13}). Finally, all $\delta^p, p\in(1,k)$ are concatenated in the channel direction to form the gradient $\delta$ of $w$. Inspired by \cite{30_grad-cam}, for the trained networks, the gradient of $w$ represents the contribution degree of each value in $w$ to the decoding result, that is, the gradient of the unimportant value in $w$ is low. In this paper, the encoded semantic vector  $w$ is firstly decoded by the transmitter, the loss is calculated using both the decoded $\hat{x}$ and the input $x$, then the loss is back propagated to get the gradient $\delta$ of $w$. After this, reshape $w$ into one dimension, and $y$ can be ordered according to the absolute value of $w$.

\subsubsection{Value and gradient of the semantic vector}
For the PCSC, assume that $l$ is the loss function of the network, $x$ is the input data, and $\hat{x}$ is the recovered data. For the trained model, $l(x,\hat{x},\gamma)$ is locally optimal and close to 0, where $\gamma$ is the network weight. If setting a value $w_{ijk}^p$ in $w$ to zero, the loss value will become $l(x,\hat{x},\gamma|w_{ijk}^p=0)$ which will be larger if $w_{ijk}^p$ is more critical. Besides, large loss values will affect the effect of the point cloud reconstruction. The square of the difference $D$ between $l(x,\hat{x},\gamma)$ and $l(x,\hat{x},\gamma|w_{ijk}^p=0)$ can be described as follows:
\begin{equation}
    \label{eq14}
    D={\left[l(x,\hat{x},\gamma)-l(x,\hat{x},\gamma|w_{ijk}^p=0) \right ]}^{2}.
\end{equation}
The first-order Taylor expansion is used in Eq. (\ref{eq14}), and the result is as follows:
\begin{equation}
    \label{15}
    D=\left [\frac{\partial l(x,\hat{x},\gamma)}{\partial w_{ijk}^p}w_{ijk}^p \right ]^2.
\end{equation}
It can be seen that $D$ is related to both the value and the gradient of $w_{ijk}^p$. To ensure good performance of the reconstructed point cloud, the discarded vector in $y$ should have smaller $D$, i.e., the absolute product of the value and gradient should be small.

\subsubsection{Realize of controllable coding rate}
The process of controllable rate coding is shown in Algorithm  \textbf{\ref{alg:cap}}. One of the above three methods can be selected in the rate allocation module to rank the importance of $y$. Then a certain number of vectors can be discarded according to the required coding rate to achieve the specified CBR. 

\begin{figure*}[b]
	\centering
	\subfloat[]{\includegraphics[width=\linewidth]{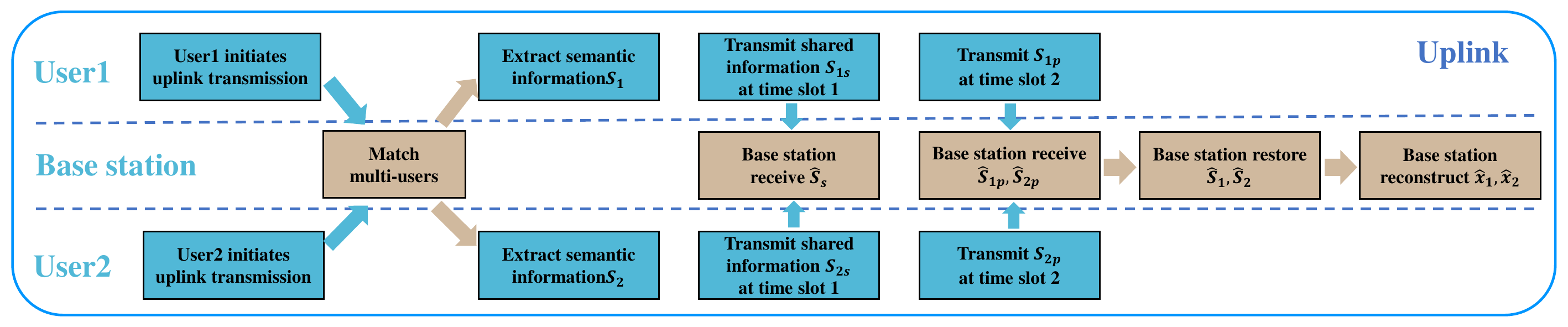}\label{MDMA_protocola}}\
    \subfloat[]{\includegraphics[width=\linewidth]{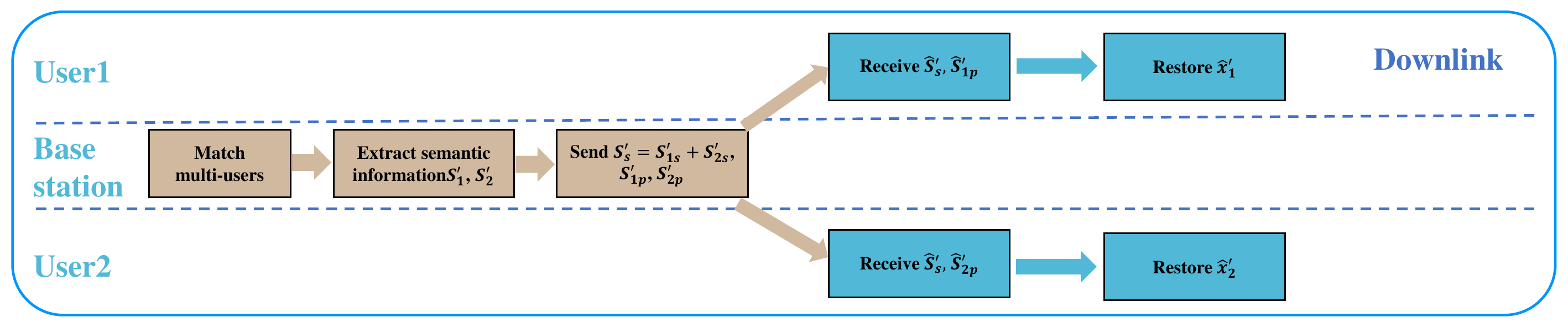}\label{MDMA_protocolb}}\
	\caption{The protocol of M-PCSC. (a): the uplink protocol of M-PCSC.  (b) the downlink protocol of M-PCSC.}
    \label{MDMA_protocol}
\end{figure*}

\subsection{Non-orthogonal Model Division Multiple Access for PCSC}
\subsubsection{M-PCSC Uplink and Downlink Design}
The uplink design of the M-PCSC is shown in Fig. \ref{MDMA_protocol}\subref{MDMA_protocola}. Considering a two-user access system, first, the base station matches two users who initiated point cloud uplink transmission instructions. Each user uses the joint source-channel encoder to encode the point cloud and transmits the shared and personalized information, respectively. The base station finally reconstructs the user’s transmitted point cloud semantic information.
The downlink design of the M-PCSC is shown in Fig. \ref{MDMA_protocol}\subref{MDMA_protocolb}. First, the base station searches for target users. Then the base station uses the joint source-channel encoder to encode the point cloud and sends the shared information and corresponding personalized information to the users. Finally, the users restore the point cloud.
In this paper, the semantic overlap rate (Sor) defined in \cite{22_MDMA} represents the resource reuse rate in the time and frequency domains. The definition of Sor is as follows:
\begin{equation}
    \label{17}
    Sor=\frac{\left | T_1\bigcap T_2\right |}{\left | T_1\right |+\left | T_2\right|},
\end{equation}
or:
\begin{equation}
    \label{18}
    Sor=\frac{\left | B_1\bigcap B_2\right |}{\left | B_1\right |+\left | B_2\right|}.
\end{equation}
Assuming that $T_i$ or $B_i$ is the time or bandwidth resource occupied by user $i (1\leq i\leq 2)$, $\left | T_i\right |$ and $\left | B_i\right |$ represent the corresponding time or bandwidth overhead. 
\par In MDMA, the user's shared information is superimposed and transmitted only once. Therefore, users' information can be transmitted with smaller bandwidth. Literature \cite{22_MDMA} presents a new metric, feasibility, denoted as $F$, to represent the service capability that the channel can provide for multiple access systems:
\begin{equation}
    \label{19}
    F=\frac{R_c}{R_s},
\end{equation}
where $R_c$ is the channel transmission rate, and $R_s$ is the source coding rate. According to the description in \cite{22_MDMA}, the feasible area of non-orthogonal MDMA uplink and downlink are greater than that of NOMA.

\subsubsection{M-PCSC Performance Analyze}
Take the downlink transmission as an example. When the base station sends the point clouds ${x}'_1$ and ${x}'_2$ to different users, it uses Eq. (\ref{eq11}) to encode and extract the shared information ${S}'_{1s}$ and ${S}'_{2s}$. The similarity between ${S}'_{1s}$ and ${S}'_{2s}$ can be measured by the absolute difference $\sigma$:
\begin{equation}
    \label{eq28}
    \sigma=\left|S'_{1s}-S'_{2s}\right|.
\end{equation}
The smaller $\sigma$ leads to the higher similarity between the ${S}'_{1s}$ and ${S}'_{2s}$. ${S}'_{1s}$ and ${S}'_{2s}$ are superimposed into ${S}'_s$ and sent to the corresponding users. In the AWGN channel, for the received $\hat{S}'_{s}$, use Eq. (\ref{eq27}) to obtain $\hat{S}'_{1s}$ and $\hat{S}'_{2s}$. In this paper, the $f(\cdot)$ is defined as follows:
\begin{equation}
\label{eq29}
f(\hat{S}'_{s})=\frac{1}{2}\hat{S}'_{s}.
\end{equation}
The $\hat{S'}_{1s}$ and $\hat{S'}_{2s}$ can be rewritten using Eq. (\ref{eq29}):
\begin{equation}
    \label{eq30}
    \hat{S}'_{1s}=\hat{S}'_{2s}=f(\hat{S}'_{s})=\frac{1}{2}(S'_{1s}+S'_{2s}+n'_0).
\end{equation}
The difference between $S'_{1s}$ and $\hat{S'}_{1s}$, $S'_{2s}$ and $\hat{S'}_{2s}$ are as follows:
\begin{equation}
\label{eq31}
\left\{
             \begin{array}{lr}
             \Delta \hat{S}'_{1s}= \left | \hat{S}'_{1s}-S'_{1s}\right |=\frac{1}{2} \left | {S}'_{1s}-S'_{2s}-n_0\right |=\frac{1}{2} \sigma + n'_0, &  \\
             \Delta \hat{S}'_{2s}= \left | \hat{S}'_{2s}-S'_{2s}\right |=\frac{1}{2} \left | {S}'_{2s}-S'_{1s}-n_0\right |=\frac{1}{2} \sigma + n'_0. & 
             \end{array}
\right.
\end{equation}
Then the base station sends personalized information ${S}'_{1p}$ and ${S}'_{2p}$ to the corresponding users. The users merge shared information and personalized information and decode using Eq. (\ref{eq12}):
\begin{equation}
    \label{eq32}
    {\hat{x}^{'}_{i}}=SC_{\beta}^{-1}(\hat{S}^{'}_{is}+\frac{1}{2}\sigma+n'_0+n'_i+\hat{S}^{'}_{ip}), i \in \left \{1,2\right \}.
\end{equation}
For the convenience of discussion, Eq. (\ref{eq32}) is simplified as follows:
\begin{equation}
    \label{eq33}
    {\hat{x}^{'}_{i}}=SC_{\beta}^{-1}(\hat{S}^{'}_{is}+\frac{1}{2}\sigma+n+\hat{S}^{'}_{ip}), i \in \left \{1,2\right \}.
\end{equation}
PSNR D1 and PSNR D2 can be used as indicators to evaluate the point cloud reconstruction performance. The specific calculation formulas of PSNR D1 and PSNR D2 will be given in Section \uppercase\expandafter{\romannumeral4}. Here, the PSNR D1 and PSNR D2 can be represented using  abstract functions:
\begin{equation}
\label{eq34}
\left\{
             \begin{array}{lr}
             \eta_{D1}(x'_{i}, \hat{x}'_{i})=\eta_{D1}(x'_{i}, SC_{\beta}^{-1}(\hat{S}^{'}_{is}+\frac{1}{2}\sigma+n+\hat{S}^{'}_{ip})), &  \\
             \eta_{D2}(x'_{i}, \hat{x}'_{i})=\eta_{D2}(x'_{i}, SC_{\beta}^{-1}(\hat{S}^{'}_{is}+\frac{1}{2}\sigma+n+\hat{S}^{'}_{ip})), & 
             \end{array}
\right.
\end{equation}
where $i \in \left \{1, 2\right\}$. The results of $\eta_{D1}$ and $\eta_{D2}$ are related to $\sigma$, $n$, the proportions of $S_{is}$ and $S_{ip}$. The proportion of ${S}'_{is}$ and ${S}'_{ip}$ can be measured by semantic overlap rate (Sor). For $\sigma$, it is also related to Sor. When the base station chooses the shared information, the absolute difference $\sigma$ of the encoded semantic vectors between two point clouds is calculated, and the $\sigma$ is then arranged from small to large. According to the specified $Sor$, the first $L\times Sor$ of the encoded vectors are selected as the shared information ($L$ is the coding length after reshaping the encoded vector into one dimension). Fig. \ref{Sfigure1} in the appendix describes the relationship between $\sigma$ and $Sor$ in different datasets, where $\sigma$ is the value which index is $L\times Sor$ in the $\sigma$ matrix in descending order. It can be seen that with the increase of $Sor$, $\sigma$ also increases correspondingly. For this reason, Eq. (\ref{eq34}) can be rewritten as follows:
\begin{equation}
\label{eq35}
\left\{
             \begin{array}{lr}
             \eta_{D1}(x'_{i}, \hat{x}'_{i})=g(Sor, SNR), &  \\
             \eta_{D2}(x'_{i}, \hat{x}'_{i})=h(Sor, SNR). & 
             \end{array}
\right.
\end{equation}
Because $\eta_{D1}$ and $\eta_{D2}$ are both functions about Sor and SNR, and the better the reconstruction performance, the higher $\eta_{D1}$ and $\eta_{D2}$ are. Here, the $\eta_{D1}$ is taken as an example to discuss, and the discussion on $\eta_{D2}$ is the same. Literature \cite{resource_allocate} definites the semantic transmission rate (S-Rate) for text semantic transmission. The S-Rate for the point cloud semantic transmission is as follows:
\begin{equation}
    \label{eq36}
    \Gamma_i=\frac{WI}{(2-Sor)L}g(Sor, SNR), i\in\left\{1,2\right\},
\end{equation}
where $W$ is the bandwidth of the transmission channel, $I$ is the average semantic information in a point cloud (measured in suts), $(2-Sor)\times L$ represents the coding length. The semantic spectral efficiency of the point cloud is further defined as follows:
\begin{equation}
    \label{eq37}
    \phi_i=\frac{\Gamma_i}{W}=\frac{I}{(2-Sor)L}g(Sor, SNR), i\in\left\{1,2\right\}.
\end{equation}

\begin{figure}[t]
\centering
\includegraphics[width=\linewidth]{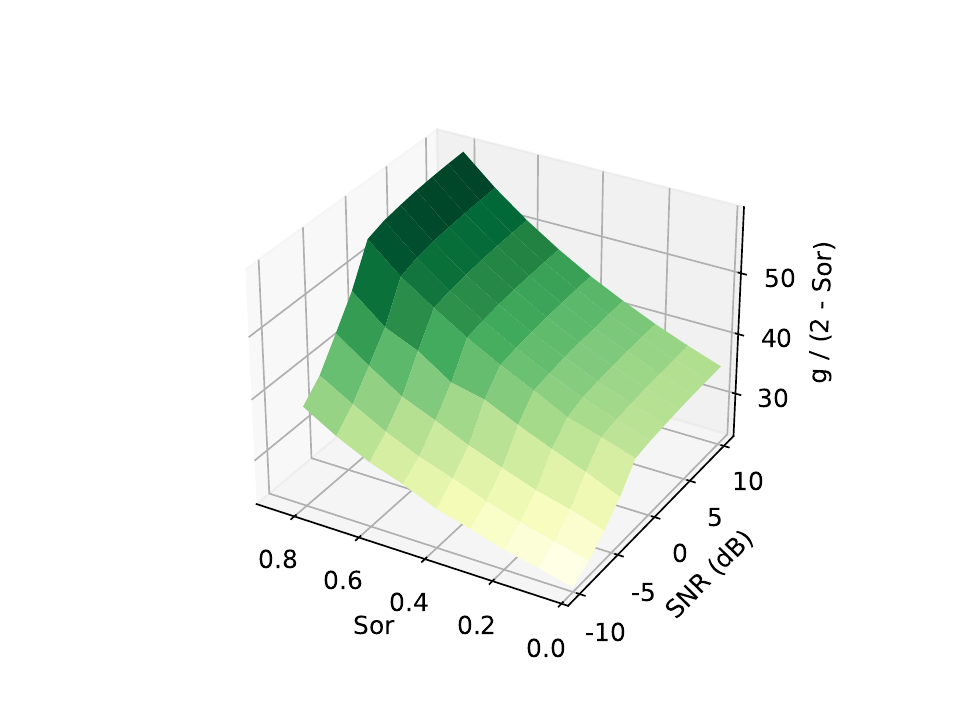}
\caption{Term $\frac{g}{2-Sor}$ in different semantic overlap rate (Sor) and SNR.}
\label{figf-sor-SNR}
\end{figure}

In this part, a semantic-aware resource allocation model based on M-PCSC is proposed to maximize $\phi_i$:
\begin{subequations}\label{eq38}
\begin{align}
\max  \quad \phi_i  \ \ \tag{\ref{eq38}}
\end{align} 
\begin{alignat}{2}
\text{s.t.}\quad &\text{$C_1$\ :\ } i \in \left\{1,2\right\},\\
    &\text{$C_2$\ :\ } 0 \leq Sor \leq 0.8, \\
    &\text{$C_3$\ :\ } g \ge g_{th}, \\
    &\text{$C_4$\ :\ } \phi_{i} \ge \phi_{th}.
\end{alignat}
\end{subequations}
According to the experiments in Section 4, M-PCSC can keep stable performance in $Sor \leq 0.8$. $C_2$ specifies the permitted range of $Sor$, $C_3$ restricts the minimum PSNR D1 by $g_{th}$, and $C_4$ reflects the minimum required $\phi_{th}$. According to \cite{resource_allocate}, term $\frac{I}{L}$ depends on the source type, which is a constant for a particular source type. For this reason, Eq. (\ref{eq38}) can be rewritten as:
\begin{subequations}\label{eq39}
\begin{align}
\max  \quad \frac{g(Sor, SNR)}{2-Sor}  \ \ \tag{\ref{eq39}}
\end{align} 
\begin{alignat}{2}
\text{s.t.}\quad &\text{$C_1$\ :\ } i \in \left\{1,2\right\},\\
    &\text{$C_2$\ :\ } 0 \leq Sor \leq 0.8, \\
    &\text{$C_3$\ :\ } g \ge g_{th}, \\
    &\text{$C_4$\ :\ } \phi_{i} \ge \phi_{th}, 
\end{alignat}
\end{subequations}
Eq. (\ref{eq39}) depends on $Sor$ and the channel conditions, so running M-PCSC on the AWGN channel to obtain the mapping between $\frac{g(Sor, SNR)}{2-Sor}$ and $(Sor, SNR)$, as shown in Fig. \ref{figf-sor-SNR}. Because $C_2$, $C_3$ and $C_4$ can only be obtained by the look-up table method, the exhausted searching method is used to solve Eq. (\ref{eq39}).

\begin{figure}[t]
\centering
\includegraphics[width=65mm]{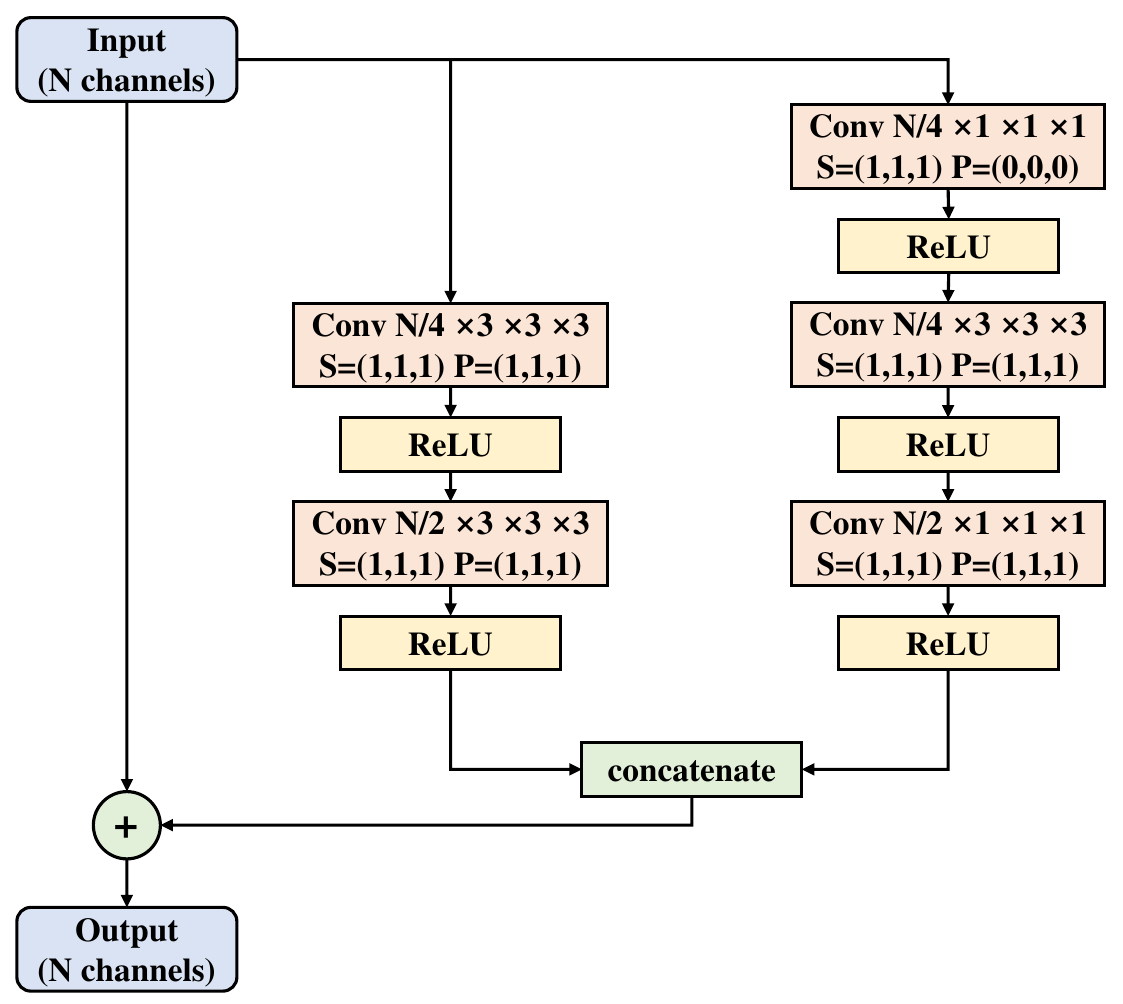}
\caption{The architecture of the Voxception-ResNet (VRN).}
\label{fig5}
\end{figure}

\begin{table}[!t]
\caption{BASE MODEL}
\label{table1}
\begin{tabular}{|c|l|}
\hline
model & \multicolumn{1}{c|}{structure} \\ \hline
\multirow{14}{*}{\begin{tabular}[c]{@{}c@{}}Joint source-channel\\ encoder\end{tabular}} & Conv3d, k=(3,3,3), s=(1,1,1), p=(1,1,1), f=16 \\ \cline{2-2} 
 & Voxception-ResNet(16) \\ \cline{2-2} 
 & Voxception-ResNet(16) \\ \cline{2-2} 
 & Voxception-ResNet(16) \\ \cline{2-2} 
 & Conv3d, k=(3,3,3), s=(2,2,2), p=(1,1,1), f=32 \\ \cline{2-2} 
 & Voxception-ResNet(32) \\ \cline{2-2} 
 & Voxception-ResNet(32) \\ \cline{2-2} 
 & Voxception-ResNet(32) \\ \cline{2-2} 
 & Conv3d, k=(3,3,3), s=(2,2,2), p=(1,1,1), f=64 \\ \cline{2-2} 
 & Voxception-ResNet(64) \\ \cline{2-2} 
 & Voxception-ResNet(64) \\ \cline{2-2} 
 & Voxception-ResNet(64) \\ \cline{2-2} 
 & Conv3d, k=(3,3,3), s=(1,1,1), p=(1,1,1), f=16 \\ \cline{2-2} 
 & Conv3d, k=(3,3,3), s=(1,1,1), p=(1,1,1), f=4 \\ \hline
\multicolumn{1}{|l|}{\multirow{14}{*}{\begin{tabular}[c]{@{}c@{}}Joint channel-source\\ decoder\end{tabular}}} & Conv3d, k=(3,3,3), s=(1,1,1), p=(1,1,1), f=16 \\ \cline{2-2} 
\multicolumn{1}{|l|}{} & Conv3d, k=(3,3,3), s=(1,1,1), p=(1,1,1), f=64 \\ \cline{2-2} 
\multicolumn{1}{|l|}{} & Voxception-ResNet(64) \\ \cline{2-2} 
\multicolumn{1}{|l|}{} & Voxception-ResNet(64) \\ \cline{2-2} 
\multicolumn{1}{|l|}{} & Voxception-ResNet(64) \\ \cline{2-2} 
\multicolumn{1}{|l|}{} & ConvT, k=(3,3,3), s=(2,2,2), p=(1,1,1), f=32 \\ \cline{2-2} 
\multicolumn{1}{|l|}{} & Voxception-ResNet(32) \\ \cline{2-2} 
\multicolumn{1}{|l|}{} & Voxception-ResNet(32) \\ \cline{2-2} 
\multicolumn{1}{|l|}{} & Voxception-ResNet(32) \\ \cline{2-2} 
\multicolumn{1}{|l|}{} & ConvT, k=(3,3,3), s=(2,2,2), p=(1,1,1), f=16 \\ \cline{2-2} 
\multicolumn{1}{|l|}{} & Voxception-ResNet(16) \\ \cline{2-2} 
\multicolumn{1}{|l|}{} & Voxception-ResNet(16) \\ \cline{2-2} 
\multicolumn{1}{|l|}{} & Voxception-ResNet(16) \\ \cline{2-2} 
\multicolumn{1}{|l|}{} & Conv3d, k=(3,3,3), s=(1,1,1), p=(1,1,1), f=1 \\ \hline
\end{tabular}
\end{table}

\section{Experiments}
In this section, the experiment of PCSC and M-PCSC will be introduced in detail. First, the experiment settings, including the datasets, the relevant training settings, the structure of the network, and the baseline, are presented. Then, the controllable rate performance, the compression performance, the transmission performance, and the M-PCSC performance will be described.

\begin{figure*}[b]
\centering
\includegraphics[height=85mm]{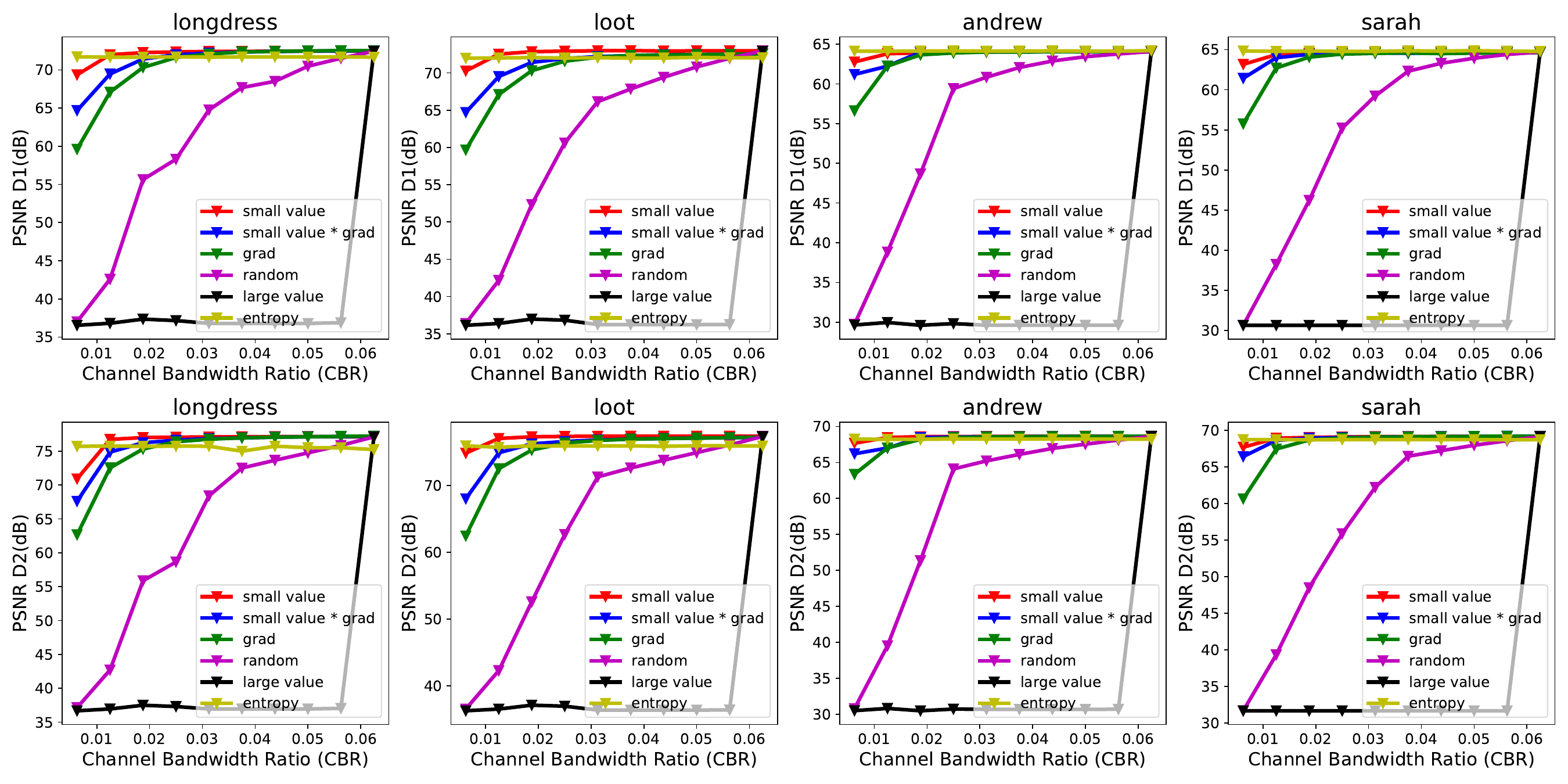}
\caption{Under AWGN channel with SNR=10dB, the controllable rate performance of PCSC in longdress, loot, andrew, and sarah: (up) D1 based PSNR, (down), D2 based PSNR.}
\label{fig6}
\end{figure*}

\begin{figure*}[t]
\centering
\includegraphics[height=85mm]{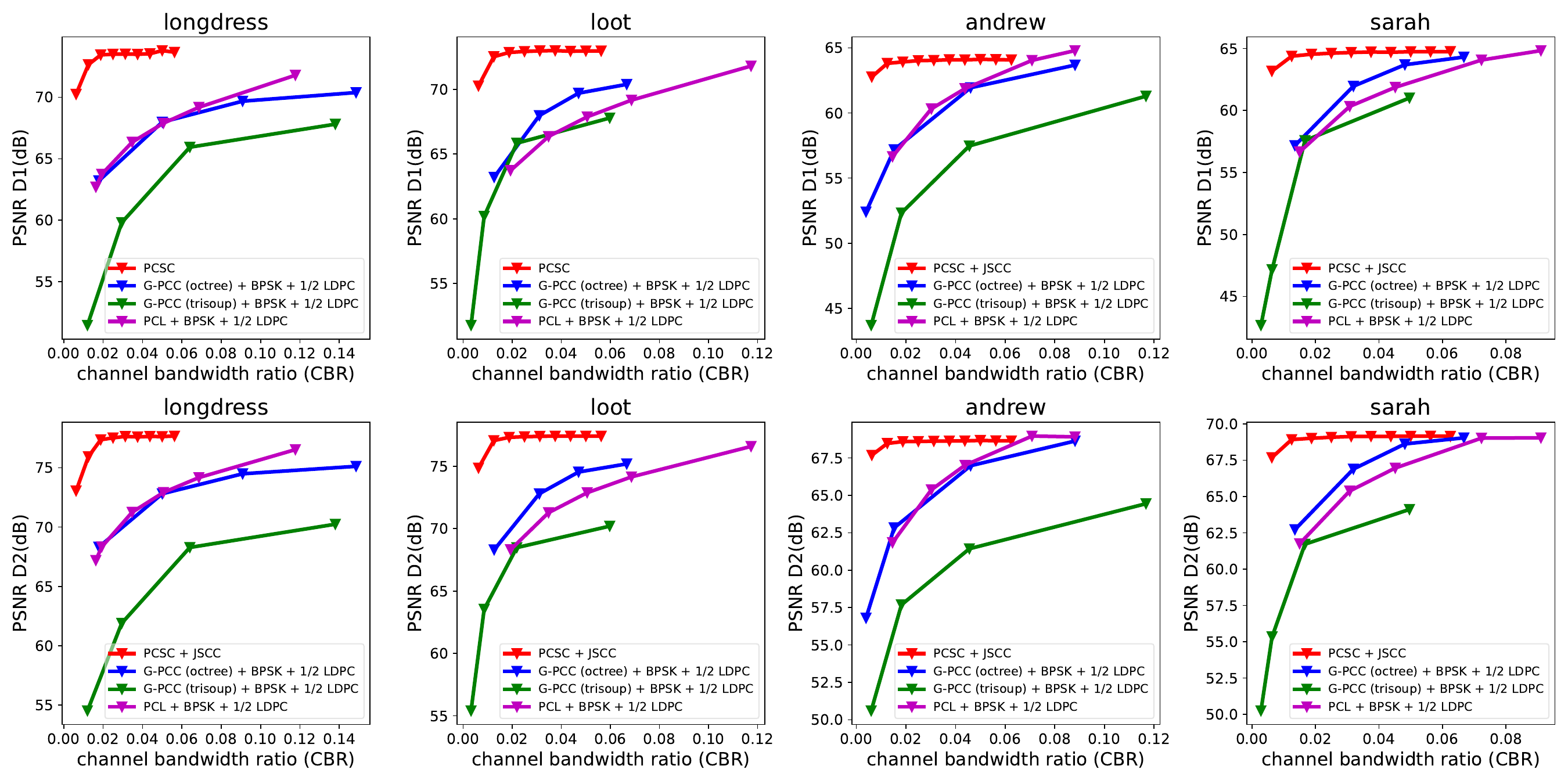}
\caption{Performance of the reconstructed point cloud in different communication systems under AWGN channel with SNR=10dB: PCSC, G-PCC(octree) + BPSK + 1/2 LDPC, G-PCC(trisoup) + BPSK + 1/2 LDPC, PCL + BPSK + 1/2 LDPC.}
\label{fig7}
\end{figure*}

\subsection{Experiments Setup}
\subsubsection{Training datasets}
About 8500 3D models from ShapeNet \cite{28_shapenet} are randomly selected for training, including 55 kinds of common objects, such as tables, chairs, cars, lamps, and so on. The point clouds from the mesh models are generated by randomly sampling points on the surface of the mesh models. Then the point clouds are voxelized into an occupied space of $256\times256\times256$. The voxelized point clouds are separated into non-overlapping cubes with the size of $64\times64\times64$. Non-overlapping cubes are randomly collected from each voxelized point cloud. A total of 180,000 cubes are used in the training. For the setting of the loss function, $\zeta=3$, and the learning rate is $10^{-5}$. The Adam optimizer \cite{29_adam} is used for iterative optimization.

\subsubsection{Testing datasets}
Two entire bodies, longdress and loot, with smooth surfaces and complete object shapes, are selected from 8i Voxelized Full Bodies (8iVFB) \cite{88i}. Another two upper bodies, Andrew and Sarah, with noisy and incomplete surfaces, are chosen from Microsoft Voxelized Upper Bodies (MVUB) \cite{9MVUB}.
\subsubsection{Baseline}
To compare the performance of PCSC under different channel bandwidth ratios (CBRs), this paper compares the proposed controllable coding rate methods with the entropy method proposed in \cite{mdvsc}. To compare the performance of different point cloud communication systems under different CBRs in the AWGN channel, compare the G-PCC \cite{35G-PCC} and PCL \cite{4PCL} using 1/2LDPC in BPSK, and the PCSC. The SNR is set to 10dB. There are two methods for G-PCC: G-PCC (octree) and G-PCC (trisoup). Regarding transmission, compare the difference between PCSC with joint source-channel coding (JSCC) and 1/2 LDPC channel coding rate in 16QAM and QPSK under different SNRs.

\begin{figure}[!t]
\centering
\includegraphics[width=\linewidth]{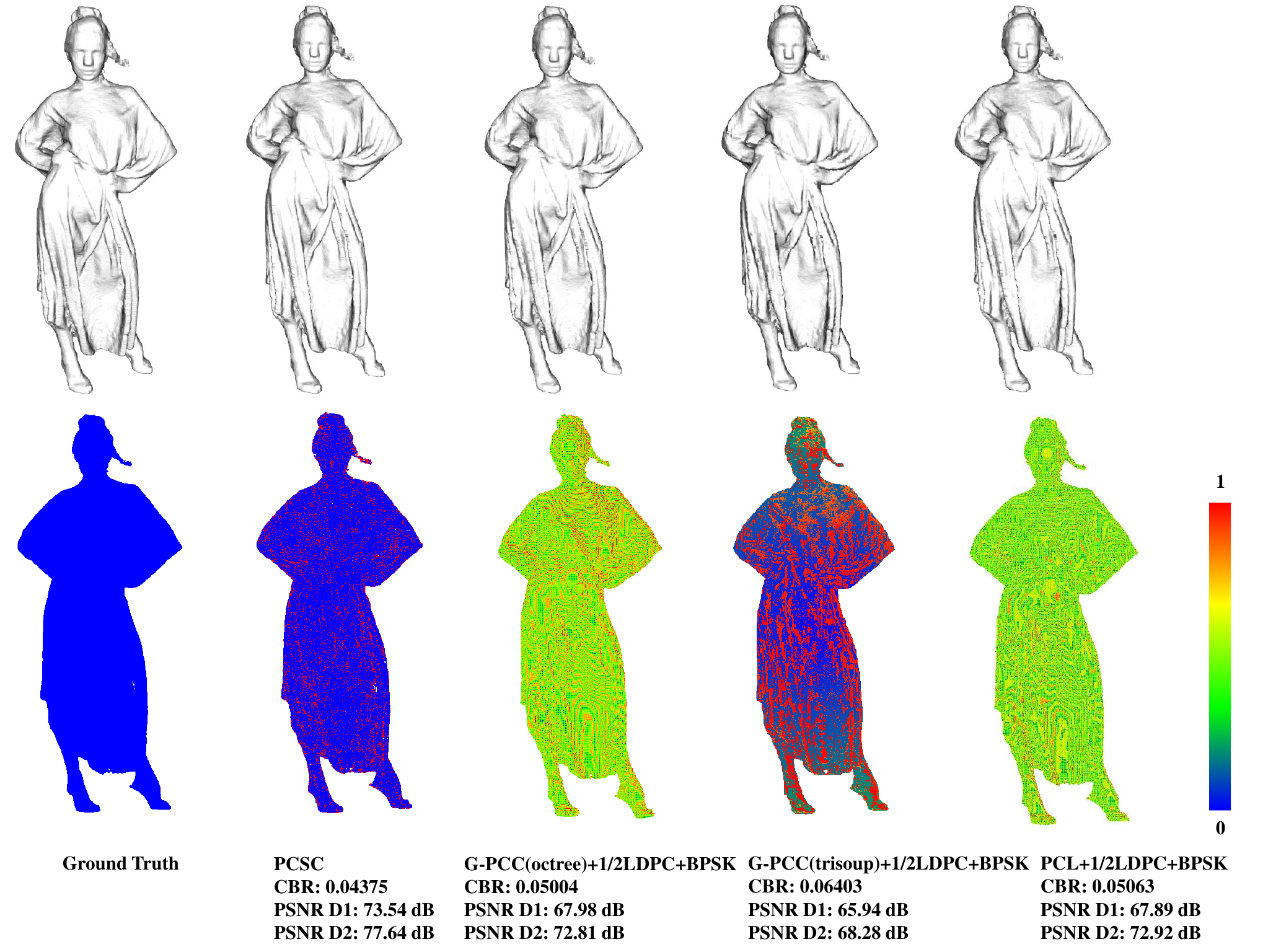}
\caption{Visual compression of longdress under AWGN channel with SNR=10dB, for ground truth, PCSC, G-PCC(octree) + BPSK + 1/2 LDPC, G-PCC(trisoup) + BPSK + 1/2 LDPC, PCL + BPSK + 1/2 LDPC. The error map (down) between the ground truth and the reconstructed point cloud is also plotted.}
\label{fig8}
\end{figure}

\subsubsection{Evaluating indicator}
In this paper, PSNR D1 and PSNR D2 \cite{31_PSNR} are used as evaluating indicators. PSNR D1 is the mean square error of point-to-point (c2c) distances in the original and reconstructed point clouds. PSNR D2 is the mean square error of point-to-plane (c2p) in the original and reconstructed point clouds.
To obtain the point-to-point error, for each point $a_j$ in the original point cloud, the nearest neighbor method is used to locate its corresponding point $b_i$ in the reconstructed point cloud. The connection $a_j$ and $b_i$ form an error vector $E(i,j)$. The length of this error vector will lead to a point-to-point error. The calculation formula is as follows:
\begin{equation}
    \label{eq22}
    e_{c2c}^{A,B}=\frac{1}{N_A}\sum {\left \| E(i,j)\right \|}^2,
\end{equation}
where A and B mean the original and reconstructed point cloud, respectively. $N_A$ is the number of points in the original point cloud.
\par To obtain the error from point to surface, the error vector $E(i,j)$ is projected along the normal vector direction $N_j$ of the original point cloud $a_j$, and a new error vector $E(i, j)$ is obtained. The error calculation formula of point to surface (c2p) is as follows:
\begin{equation}
    \label{eq23}
    e_{c2p}^{A,B}=\frac{1}{N_A}\sum {( E(i,j)\cdot N_j)}^2,
\end{equation}
where A and B mean the original and the reconstructed point cloud, respectively. $N_A$ is the number of points in the original point cloud.
\par Eq. (\ref{eq22}) and Eq. (\ref{eq23}) are measured by the mean square error (MSE). However, the MSE between multiple-point clouds is difficult to understand. To facilitate understanding, MSE is converted to a PSNR using the following equation:
\begin{equation}
    \label{24}
    PSNR_{A,B}=10\log_{10}\frac{p^2}{e_{A,B}},
\end{equation}
where $p$ is the peak value of the signal. In this paper, $p=\sqrt{3} \times (2^b-1)$, and $b$ is the precision of the point cloud. The precision of the longdress and the loot is 10, and 9 for the andrew and the sarah.

\subsubsection{Model structure}
The structure of the Voxception-ResNet used in this paper is shown in Fig. \ref{fig5}, and the basic structures of the encoder and the decoder are shown in Table \ref{table1}, where k, s, p, f mean the kernel, stride, padding, feature, respectively. The joint source-channel encoder extracts semantic features, compresses redundant information, and codes channels. The decoder structure, which is symmetrical to the encoder, is used for recovering the received semantic features into a point cloud.

\begin{figure*}[t]
	\centering
	\subfloat[AWGN]{\includegraphics[height=85mm]{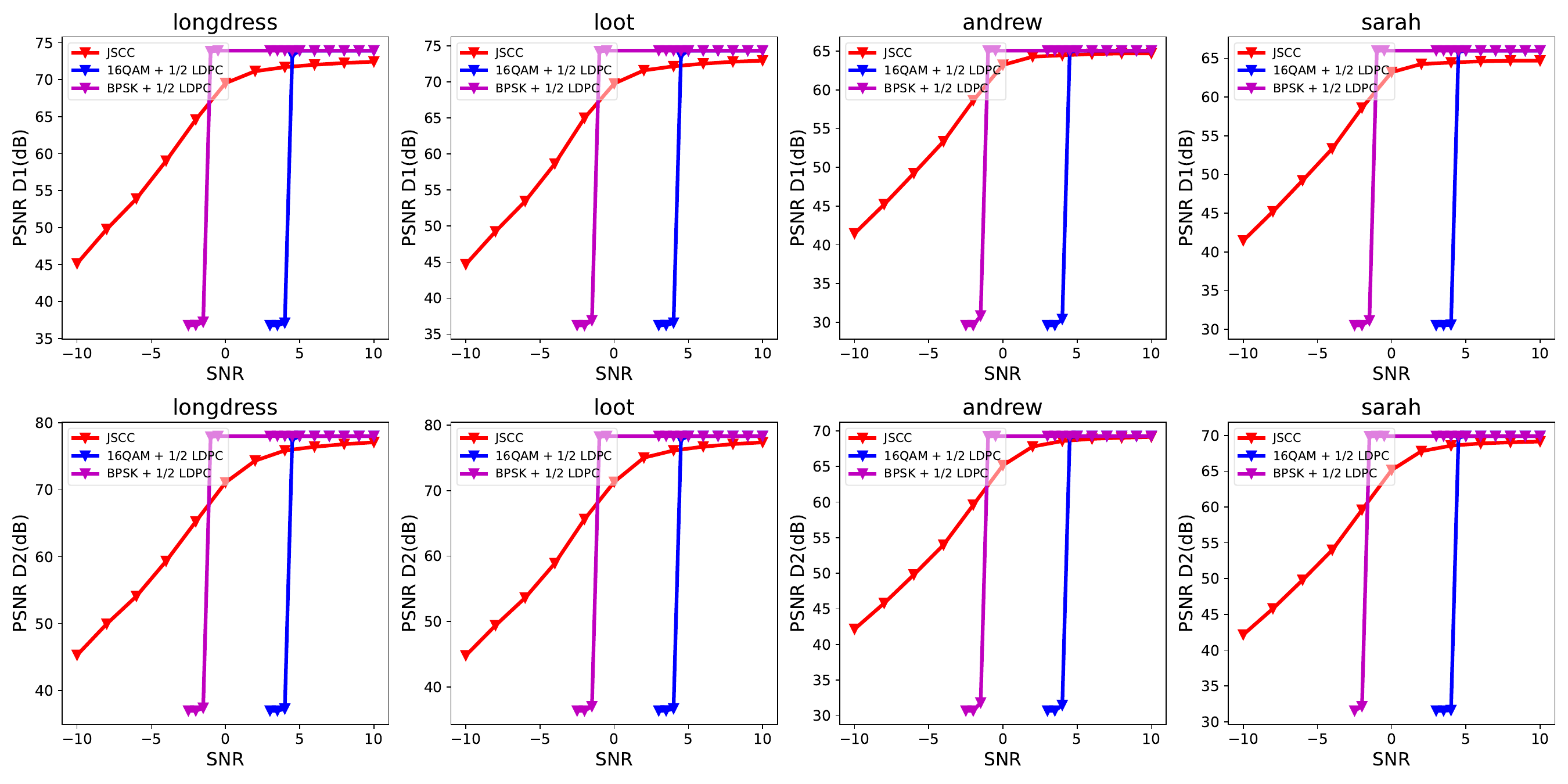}}\\
	\subfloat[Rayleigh]{\includegraphics[height=85mm]{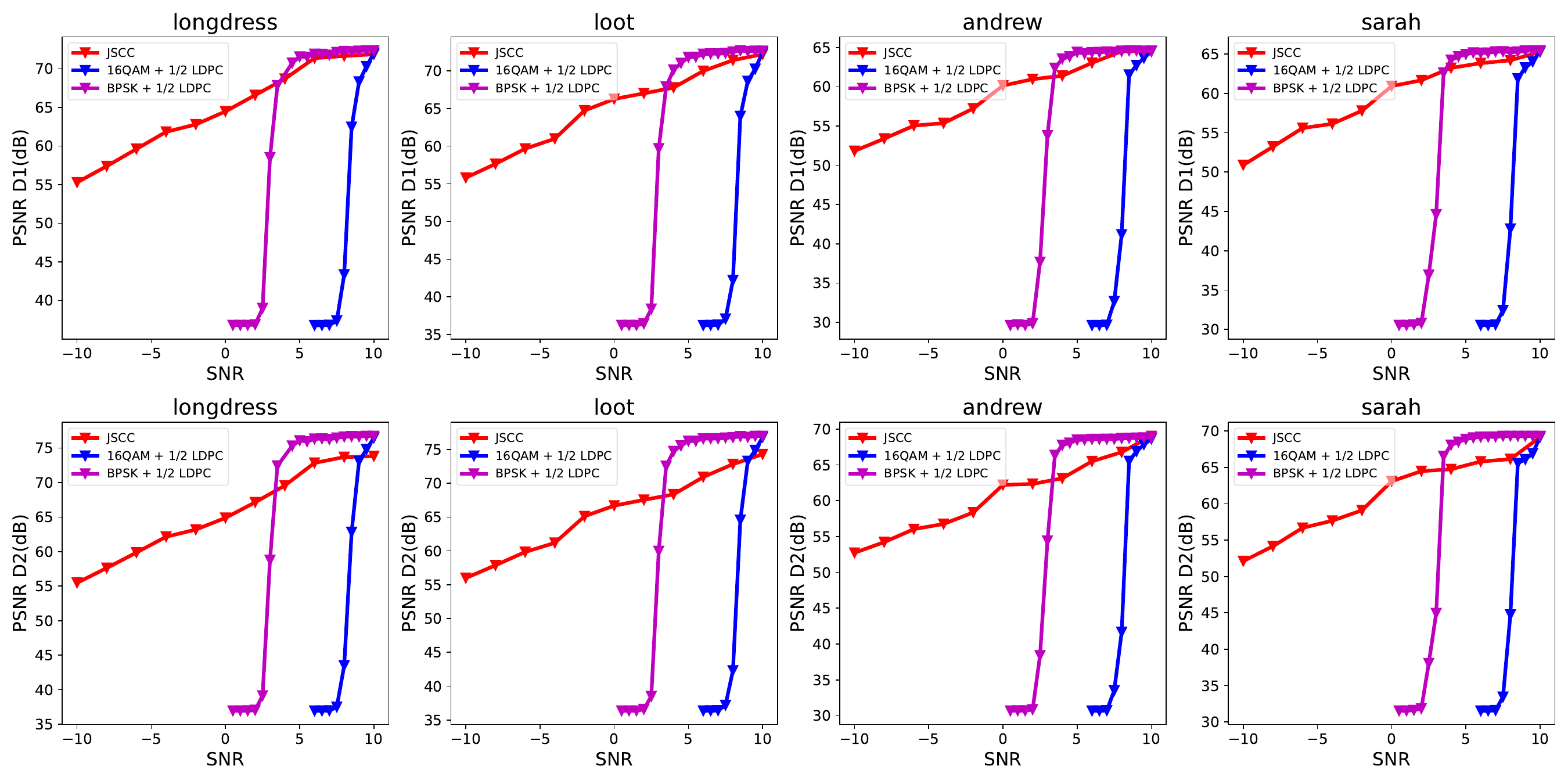}}
	\caption{Performance comparison of PCSC with JSCC, 16QAM + 1/2 LDPC, BPSK + 1/2 LDPC in AWGN channel and Rayleigh Fadding channel. The channel bandwidth ratio (CBR) is set to 0.0625.}
    \label{fig.9}
\end{figure*}

\subsection{Result Analysis}
\subsubsection{PCSC performance}
Fig. \ref{fig6} shows the PSNR D1 and PSNR D2 at different coding rates over an AWGN channel with SNR=10dB. The channel bandwidth ratio CBR=0.0625 without discarding any points, and the CBR after dropping points are less than 0.0625. In this experiment, the ratio of discarding points was set to 0–0.9, with a step space of 0.1. The value, the gradient, and the product of the gradient and value are used as the basis for ranking semantic vectors. As mentioned above, the network learns the degree of importance and the adaptation of power allocation, so the performance is still reliable after discarding some points with low power (legend is small value) but worse after discarding some points with high power (legend is large value). The performance achieved by the methods proposed in this paper, especially at low CBR, is about 25dB higher than the performance of random dropping points. However, due to the way the three methods rank the semantic vector differently, the performance of the three methods is different at low CBR. In addition, the performance of the method proposed in this paper is stable where the CBR is 0.03125 (dropping rate is 50$\%$) on longdress and loot, and stable where the CBR is 0.01875 (dropping rate is 70$\%$) on andrew and sarah. This paper's method can effectively reduce the transmission bandwidth and still perform well under the condition of reduced CBR. Besides, the performance of the method proposed in this paper (value, gradient, value × gradient) is equivalent to that of the entropy method when the CBR is larger than 0.0125 (dropping rate is 80$\%$) and is slightly lower than that of the entropy method when the CBR is less than 0.0125. The methods used in this paper do not need to add additional networks for training, only analyze the importance of the encoded semantic vectors and discard some semantically-unimportant vectors. The entropy method needs to modify the original network and retrain the network before ranking the importance of the vectors. Therefore, when the required CBR is not very small, the method proposed in this paper can be adopted, and when the required CBR is very small, the entropy method can be adopted.

\par Fig. \ref{fig7} shows the performance curve of the PSNR D1 and the PSNR D2 achieved by different communication systems. The SNR is set to 10dB. For the G-PCC (octree), the G-PCC (trisoup), and the PCL, use the noise-robust combination BPSK + 1/2 LDPC. It can be seen that PCSC has relatively stable performance under different CBR, which other methods can not achieve. Although the lowest CBR achieved by PCSC is slightly higher than that of G-PCC(trisoup), it is enough to show that PCSC can meet the existing compression performance. PCSC thinks that the symbols representing information are unequal. For PCSC, when transmitting the point cloud under different CBR, the importance of the coded semantic vectors is analyzed, and then the semantically-unimportant vectors are discarded. For other methods, it is considered that the symbols representing information are equal. When transmitting the point cloud in low CBR, the points in the input point cloud are directly reduced and then coded, likely losing some points with high importance. With the increase of CBR, more points are used for coding and transmission, and the reconstruction performance is improved. Therefore, the point cloud reconstruction performance of the PCSC is better than other methods at low CBR, and the performance of the PCSC is close to that of other methods at high CBR. Besides, the decoded point clouds and the ground truth is shown in Fig. \ref{fig8}. The error map based on the point-to-point distance between decoded point clouds and ground truth is also plotted. Compared with other methods, the method used in this paper has minor errors.

\begin{figure*}[t]
	\centering
	\subfloat[]{\includegraphics[height=42.5mm]{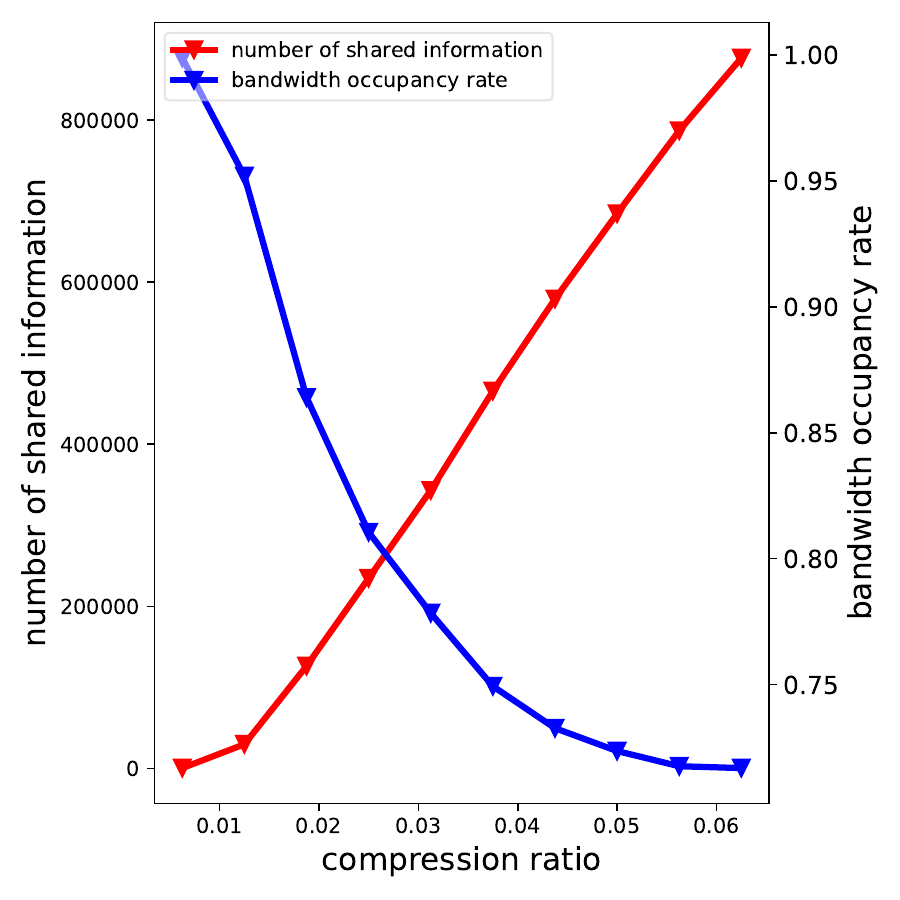}\label{10a}}\
    \subfloat[]{\includegraphics[height=42.5mm]{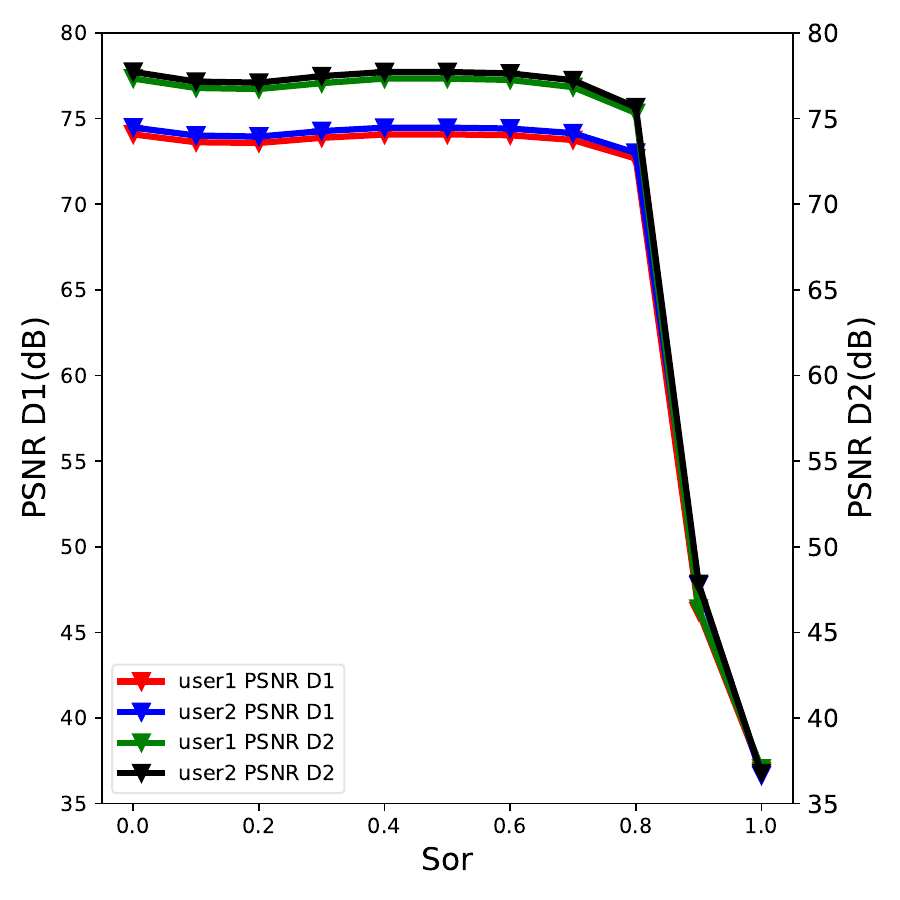}\label{10b}}\
	\subfloat[]{\includegraphics[height=42.5mm]{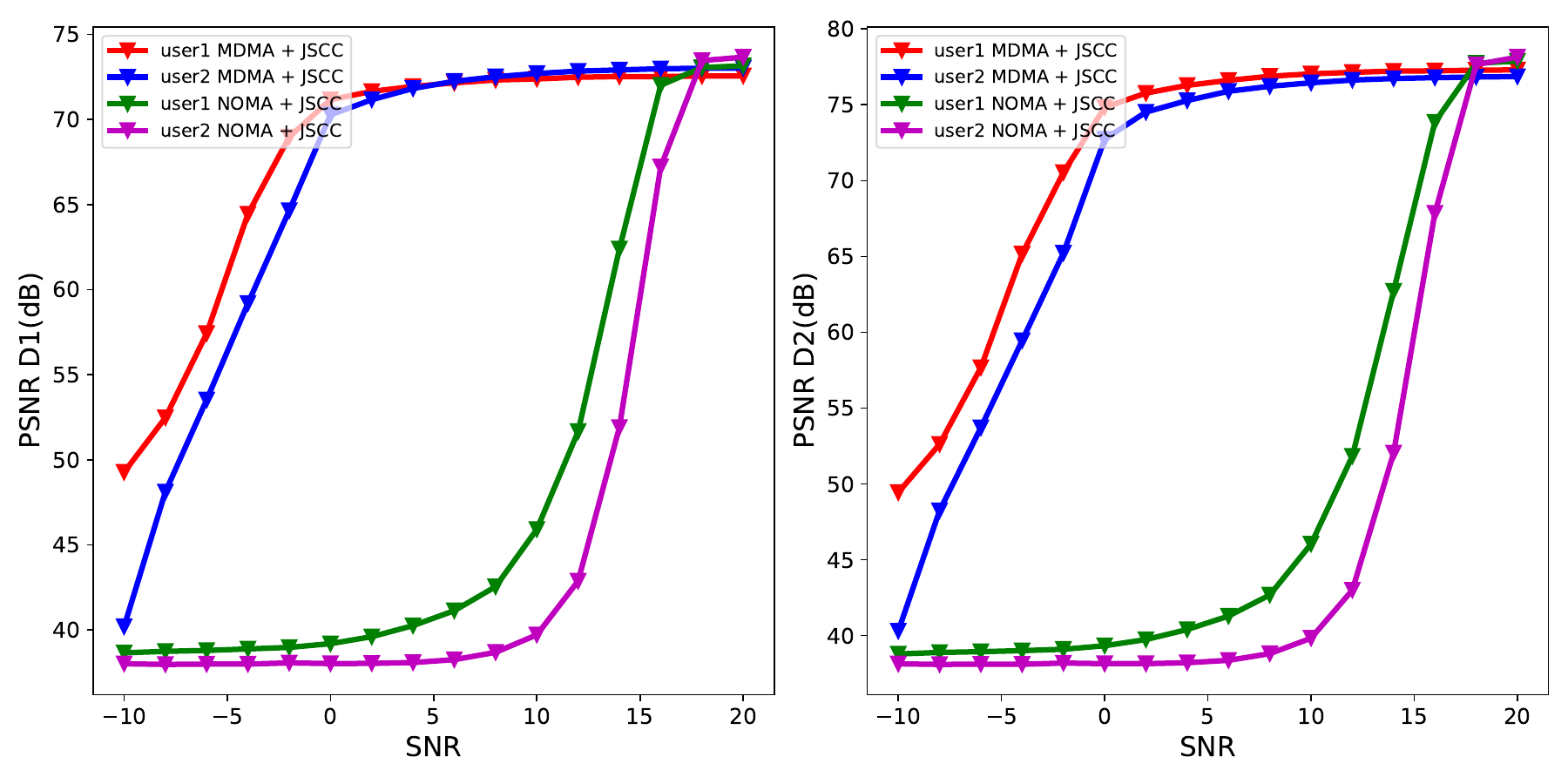}\label{10c}}
	\caption{M-PCSC performance. (a): the number of shared information and the bandwidth occupancy rate under different compression ratios.  (b) the PSNR D1 and PSNR D2 performance under different semantic overlap rates (Sor). (c) the comparison between PCSC-based non-orthogonal MDMA (M-PCSC) and NOMA.}
    \label{fig.10}
\end{figure*}

\par Fig. \ref{fig.9} shows the transmission performance under different SNRs. The PCSC is trained once under the SNR=10dB and tested under various SNRs. The CBR is set to 0.0625. This paper compares the differences between the PCSC used joint source-channel coding (JSCC) in the AWGN and Rayleigh channels and the PCSC with 1/2 LDPC + 16QAM and 1/2 LDPC + BPSK. Under the high SNR, the performance of both JSCC and LDPC coding is stable, and the performance of LDPC is slightly higher than that of JSCC. With the decrease of SNR, the performance of the PCSC using LDPC coding drops sharply, appearing as a ``cliff effect”. In the AWGN channel, the ``cliff effect" of 1/2LDPC+16QAM and 1/2LDPC+QPSK occurs at about 4.5dB and -1dB, respectively. In the Rayleigh channel, the ``cliff effect" of 1/2LDPC+16QAM and 1/2LDPC+QPSK occurs at approximately 9dB and 4.5dB, respectively. The JSCC effectively alleviates the ``cliff effect". With the reduction of SNR, the system performance does not decline sharply but slowly, which indicates that the anti-noise ability of JSCC is relatively strong. In PCSC+1/2 LDPC+16QAM and PCSC+ 1/2 LDPC+BPSK, it is considered that the bit error location is random and not related to the semantic features of the point cloud behind the bit stream. In low SNR, the reconstruction performance drops sharply once an error occurs. In PCSC+JSCC, the whole system optimizes the end-to-end transmission distortion and considers the semantic features of the point cloud. Therefore, the model can learn how to preserve semantic features from noise.

\subsection{M-PCSC performance}
First, the original point cloud is compressed using the encoder of PCSC with different compression rates. The compression ratio is set to the file size after compression divided by the original file size. When the absolute difference between two semantic vectors is less than 0.001, the two vectors are considered to be similar. As shown in Fig. \ref{fig.10}\subref{10a}, it can be seen that the number of shared information between the two point clouds is gradually reduced with the reduction of the compression rate. When that compression ratio is extremely low, the number of shared information between the two point clouds is almost 0. It can be seen that the compression is mainly for shared information. Because M-PCSC transmits shared information once, and transmits personal information separately. Therefore, more bandwidth is needed when the compression ratio is low, because the proportion of shared information between the two point clouds is small. For this reason, the bandwidth occupation rate will increase with the reduction of the compression rate when using the M-PCSC.
\par Fig. \ref{fig.10}\subref{10b} shows the performance of M-PCSC with increased semantic overlap rate (Sor) over a Gaussian channel with SNR=10dB. When Sor is set to 0 and 1, the bandwidth occupancy rate is 1 and 0.5, respectively. When Sor is less than 0.8, the recovery performance of the two users is almost stable. When the Sor reaches 0.8, the performance is close to the original recovery, which means it can save 40$\%$ bandwidth. When the Sor is larger than 0.8, the performance degrades gradually.
\par To further demonstrate the performance of M-PCSC, the performance of M-PCSC and point cloud transmitting using NOMA in the case of PCSC is compared. For a fair comparison, the Sor of the M-PCSC is set to 0.8, and the CBR is also adjusted to ensure the bandwidth occupied by the M-PCSC is consistent with that occupied by NOMA. The experimental results are shown in Fig. \ref{fig.10}\subref{10c}. It can be seen that under a high SNR, the effect of NOMA is slightly higher than that of non-orthogonal MDMA. However, under a low SNR, the advantages of non-orthogonal MDMA are much greater than those of NOMA. When the SNR is 0dB, the performance of non-orthogonal MDMA is better than that of NOMA by about 30 dB.

\section{Conclusion}
This paper proposes a new point cloud semantic communication system (PCSC) and a simple but efficient method to control the coding rate. The value and the gradient of the encoded semantic vector are taken as a basis to analyze the importance degree of the semantic vector, and a certain proportion of unimportant-semantically data is discarded to save bandwidth. In addition, a system combining PCSC and the new proposed non-orthogonal model division multiple access (MDMA) technology is proposed, named M-PCSC, which can effectively reduce the transmission bandwidth when two point clouds are transmitted simultaneously. The entire system is described as an optimization problem to minimize end-to-end transmission distortion. Relevant experimental results show that the proposed communication system can generally exceed the traditional methods, and has a large increase in PSNR D1 and PSNR D2 indicators.

\section*{Acknowledgment}
This work is supported in part by the National Key R$\&$D Program of China under Grant 2022YFB2902102.

\newpage
{\appendix[The relationship between $Sor$ and $\sigma$]
\begin{figure}[hb]
\centering \includegraphics[width=\linewidth]{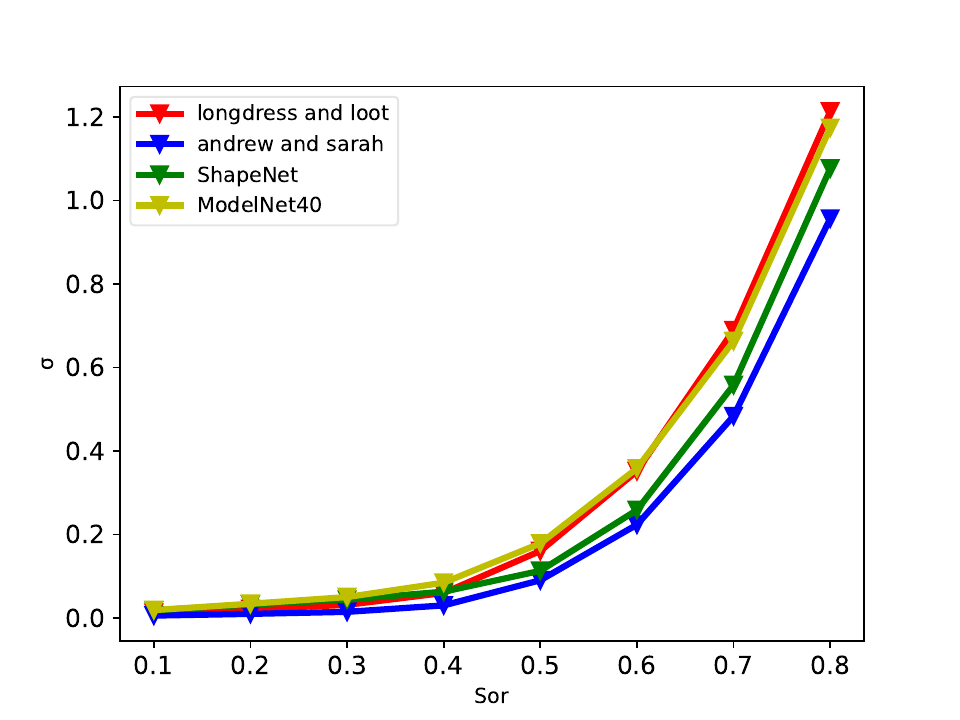}
\caption{The relationship between $Sor$ and $\sigma$ with the longdress and loot dataset, andrew and sarah dataset, ShapeNet, and ModelNet40 \cite{modelnet}.}
\label{Sfigure1}
\end{figure}

}

\bibliographystyle{IEEEtran}
\small\bibliography{reference}

\vfill

\end{document}